\documentclass[trackchanges,twocolumn]{aastex701}
\usepackage{lipsum,amsmath,placeins}
\hypersetup{linkcolor=red,citecolor=blue,filecolor=cyan,urlcolor=blue}
\graphicspath{{./}{Figures/}}

\begin{document}

\title{Formation versus Dynamical Survival: Decoding the Temperature-dependent Reversal of the Wide-binary Fraction}

\author[gname=Sujie,sname='Guo']{Sujie Guo}
\affiliation{School of Physics, Hangzhou Dianzi University, Hangzhou 310018, People's Republic of China}
\affiliation{Zhejiang Branch, National Astronomical Data Center, Hangzhou 310018, People's Republic of China}
\email{gsj@hdu.edu.cn}  

\author[orcid=0000-0003-3957-9067, gname=Xunzhou, sname='Chen']{Xunzhou Chen} 
\affiliation{School of Physics, Hangzhou Dianzi University, Hangzhou 310018, People's Republic of China}
\affiliation{Zhejiang Branch, National Astronomical Data Center, Hangzhou 310018, People's Republic of China}
\email[show]{cxz@hdu.edu.cn}

\author[orcid=0000-0001-9289-0589,gname=Haijun, sname='Tian']{Haijun Tian}
\affiliation{School of Physics, Hangzhou Dianzi University, Hangzhou 310018, People's Republic of China}
\affiliation{Zhejiang Branch, National Astronomical Data Center, Hangzhou 310018, People's Republic of China}
\affiliation{Institute of Astronomy, Cambridge University, Cambridge CB3 0HA, UK} 
\email[show]{hjtian@hdu.edu.cn}

\begin{abstract}

We analyze 52,957 high-purity main-sequence wide binaries from Gaia EDR3 and LAMOST DR9 to investigate the physical origin of the relative wide-binary fraction (WBF)--metallicity relation. In the low-\(\alpha\) population, we find a striking temperature-dependent reversal: the relative WBF decreases with [Fe/H] for cool stars (\(4000\,{\rm K}<T_{\rm eff}<5400\,{\rm K}\)) but increases with [Fe/H] for hot stars (\(6000\,{\rm K}<T_{\rm eff}<7500\,{\rm K}\))---a reversal that is entirely absent in the high-\(\alpha\) population, where the relative WBF rises monotonically with [Fe/H] at wide separations. In contrast, the relative WBF decreases with age across all three temperature subsamples, with no corresponding reversal. However, the opposing WBF–[Fe/H] trends persist within the same age bins, suggesting that neither differences in age distributions nor variations in age–[Fe/H] relations can fully account for the observed reversal. We suggest that the negative [Fe/H] dependence among cool stars may preserve a formation-related imprint through subsequent dynamical widening, whereas the positive dependence among hot stars may reflect mass-dependent multiplicity and variations in birth environments. These findings provide new constraints on wide-binary formation and subsequent dynamical evolution.

\end{abstract}


%Section
\section{Introduction}

Binary and multiple stellar systems are fundamental outcomes of star formation. Their binary fractions, mass-ratio distributions, and orbital properties provide important constraints on theories of stellar formation, evolution, and Galactic dynamics \citep{Duchne2013,Moe2017,Chen2024}. Wide binaries, commonly defined as spatially resolved co-moving stellar pairs with projected separations ($s$) of $\sim 10^2$--$10^5{\rm AU}$, are particularly valuable systems. The components are generally coeval and chemically homogeneous, making them useful benchmarks for stellar-abundance and stellar-age measurements \citep{Makarov2008,Kraus2009,Chanam2012,Andrews2017,Andrews2019,Hawkins2020,Shariat2026}. Owing to their large orbital separations and weak mutual interactions, the components can generally be treated as independently evolving stars. Their weak binding also makes wide binaries sensitive tracers of Galactic dynamical perturbations, including Galactic tides \citep{CorreaOtto2017,Hamilton2022,Modak2023}, field star encounters and impulsive perturbers \citep{Jiang2010,Kaib2013,Kaib2014,Hamilton2024,Yavetz2025,Caputo2026}, as well as dark matter substructure \citep{Yoo2004,Ramirez2023,Shariat2025}.

Metallicity regulates gas cooling, radiative feedback, protostellar-disk stability, and fragmentation efficiency. Different formation channels may therefore imprint distinct signatures on the relation between the wide-binary fraction (WBF) and metallicity \citep{Machida2009,Bate2014,Moe2019}. The dependence of the WBF on metallicity is a key diagnostic of binary formation and evolution. Studies of close binaries have shown that the binary fraction decreases with increasing [Fe/H] \citep{Moe2019}. For wide binaries, however, the nature of this relation remains unsettled. The WBF of very wide binaries has been suggested to depend only weakly on [Fe/H] \citep{ZapateroOsorio2004}; \citet{El-Badry2019} found that the relation becomes significantly anti-correlated at $s\lesssim200\,{\rm AU}$; \citet{Hwang2021} reported that the WBF peaks near solar metallicity for binaries with $10^3\lesssim s\lesssim10^4\,{\rm AU}$; and \citet{Niu2022} further demonstrated that the relation depends on effective temperature ($T_{\rm eff}$) and population membership. At very low metallicities, \citet{Shariat2025} found a WBF comparable to the solar-neighborhood value at $[{\rm Fe/H}]\approx-2.5$, whereas \citet{Lodieu2025} reported a lower binary fraction for stars with $[{\rm Fe/H}]<-1.5$.  These results indicate that the WBF-[Fe/H] relation is jointly shaped by formation physics, $s$, population, and subsequent dynamical evolution--a complexity that {we investigate by controlling for $s$, $T_{\rm eff}$, and stellar population, while examining the roles of stellar age and Galactic guiding radius.}

The advent of Gaia and large-scale spectroscopic surveys has provided an unprecedented opportunity to address this problem. Before Gaia, studies of wide binaries were limited by sample size, contamination, and distance reach, making it difficult to distinguish genuine physical pairs from chance alignments on the sky \citep{Lpine2007,Raghavan2010,Tokovinin2014}. {Gaia astrometry has since enabled the construction of large, high-purity wide-binary catalogs \citep{Gaia2016,Gaia2021,ElBadry2021,ElBadry2024,Shariat2026}.} {In parallel, LAMOST has provided spectroscopy for millions of Galactic stars \citep{Cui2012,Luo2015}, and the DD-Payne framework yields homogeneous stellar parameters and multi-element abundances for large samples \citep{Xiang2019,Ting2019,Zhang2025}.} {Together, Gaia astrometry and LAMOST spectroscopy provide the kinematic and chemical information needed for a large-sample study of wide binaries.}

In this work, we combine Gaia and LAMOST data to investigate the metallicity dependence of the {relative WBF} among main-sequence binaries across different ranges of $s$, $T_{\rm eff}$ and stellar populations. The paper is organized as follows. Section~\ref{Data and Methods} describes the data, sample selection, and methodology. Section~\ref{results} presents the {relative-WBF}-[Fe/H] relations and {investigates the physical origin of the temperature-dependent reversal in the low-$\alpha$ population.} Section~\ref{summary} summarizes the main findings.

\section{Data and Methods}
\label{Data and Methods}

In this section, we combine Gaia with LAMOST to construct a high-purity MS--MS wide-binary sample and stratify it by $s$, $T_{\mathrm{eff}}$, and population. We define the {relative WBF} using matched control samples and assess distance-dependent incompleteness with the twin-binary test.

\subsection{Wide--binary sample construction and quality selection}

We construct the wide-binary sample by cross-matching the Gaia EDR3 wide-binary catalog of \citet{ElBadry2021} with the LAMOST DR9 DD-Payne value-added catalog \citep{Zhang2025}. The wide-binary catalog contains approximately 1.3 million resolved wide binaries within 1 kpc of the Sun, with $s$ ranging from a few AU to $\sim 1$ pc. The LAMOST DR9 DD-Payne catalog provides homogenized stellar parameters and 22 elemental abundances for FGK stars. The underlying model was trained using high-resolution spectroscopic surveys, including GALAH and APOGEE, within the data-driven Payne (DD-Payne) framework. In this catalog, [Fe/H] has been corrected for non-local thermodynamic equilibrium (non-LTE) effects, $T_{\rm eff}$ has been calibrated to the infrared flux method scale, and $\log g$ has been validated against asteroseismic constraints.  

To obtain a high-purity astrometric binary sample, we retain only MS--MS pairs and require a chance-alignment probability of $R<0.1$, corresponding nominally to a physical-association probability greater than 90\%. We then apply quality cuts to ensure reliable stellar parameters and chemical abundances: ${\rm SNR}_G > 20$, ${\rm ChisqFlag\_LAM\_APO} < 3$, ${\rm ChisqFlag\_LAM\_GAL} < 3$, and valid abundance flags for Mg, Si, Ca, and Ti (i.e. ${\rm Flag}_{X/{\rm Fe}}=0$ for all four elements), with [$\alpha$/Fe] computed as their inverse-variance-weighted mean following \citet{Xiang2019}. In the crossmatched sample, LAMOST spectra are available for 59,423 primaries and 9,789 secondaries, where the primary and secondary labels follow the Gaia $G$-band magnitude definition of \citet{ElBadry2021}. After merging the matched components, we obtain 61,983 wide-binary systems, {including 7,229 with LAMOST measurements for both components.} {Wide-binary components are generally expected to be co-natal and chemically homogeneous \citep{Makarov2008,Kraus2009,Andrews2017,Andrews2019,Hawkins2020}. We therefore adopt the mean chemical abundances when valid measurements are available for both components.} For the temperature-based subsamples, we adopt the primary-component $T_{\rm eff}$ when available and otherwise use that of the secondary.

\begin{figure*}
\centering
\includegraphics[width=0.98\textwidth]{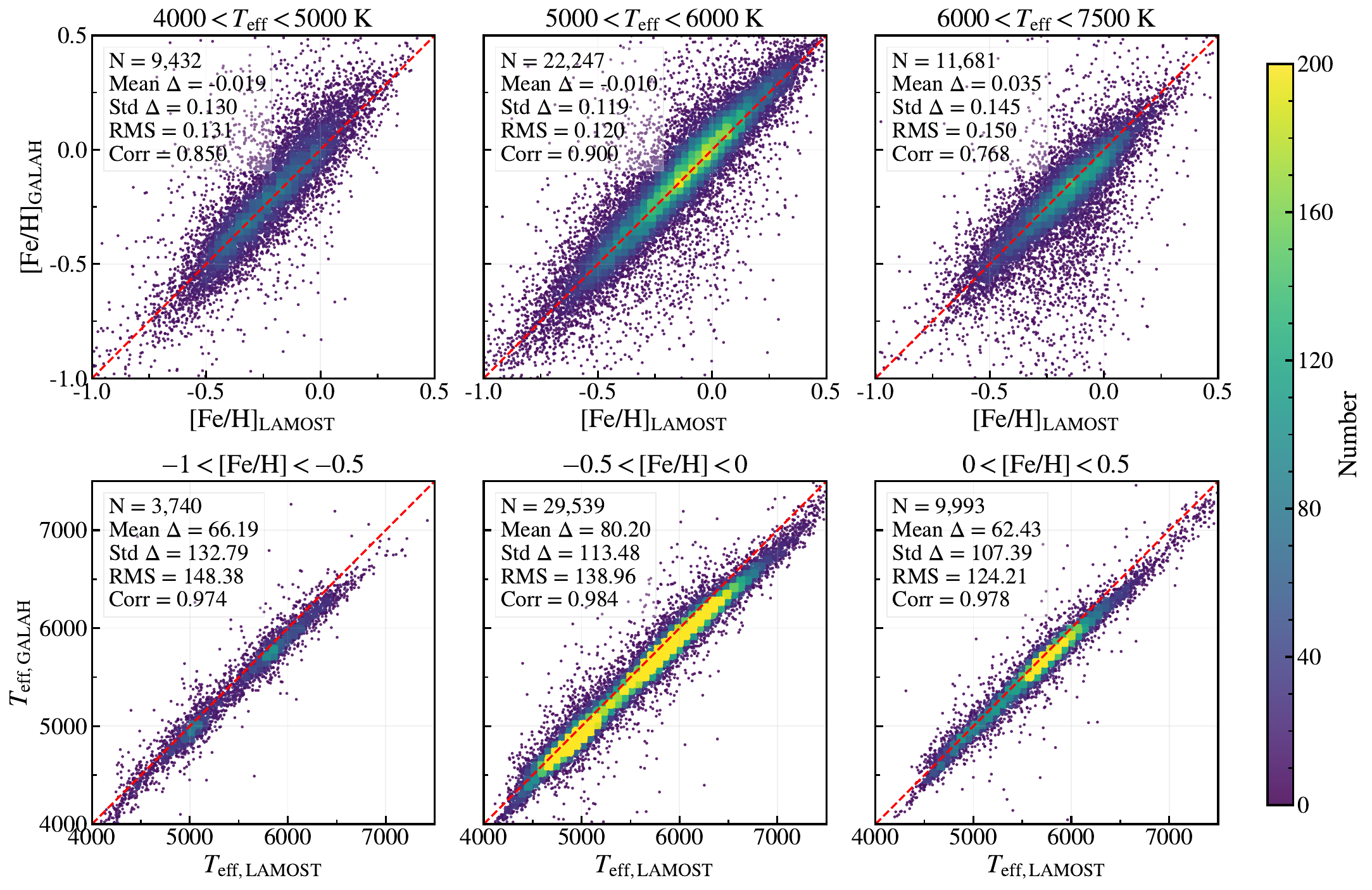}
\caption{Comparison of stellar parameters between the LAMOST DR9 (DD-Payne) and GALAH DR3 catalogs. The upper panels compare [Fe/H] in three $T_{\rm eff}$ intervals, while the lower panels compare $T_{\rm eff}$ in three [Fe/H] intervals. The red dashed lines indicate the one-to-one relations. The color bars show the number of data points. The mean offset, standard deviation, RMS scatter, and correlation coefficient are listed in each panel.}
\label{fig:lamost_galah}
\end{figure*}

An unbiased metallicity ([Fe/H]) scale is essential for accurately characterizing the dependence of the WBF on [Fe/H]. Large spectroscopic surveys often exhibit $T_{\rm eff}$-dependent systematic offsets in [Fe/H], especially for cool stars. Given that our analysis relies on the LAMOST catalog, we crossmatch it with GALAH DR3 to test for $T_{\rm eff}$-dependent systematics in [Fe/H]. Figure~\ref{fig:lamost_galah} shows consistent agreement across different $T_{\rm eff}$ and [Fe/H] ranges, with no evident parameter-dependent offset. 

\begin{figure}
\centering
\includegraphics[width=\linewidth]{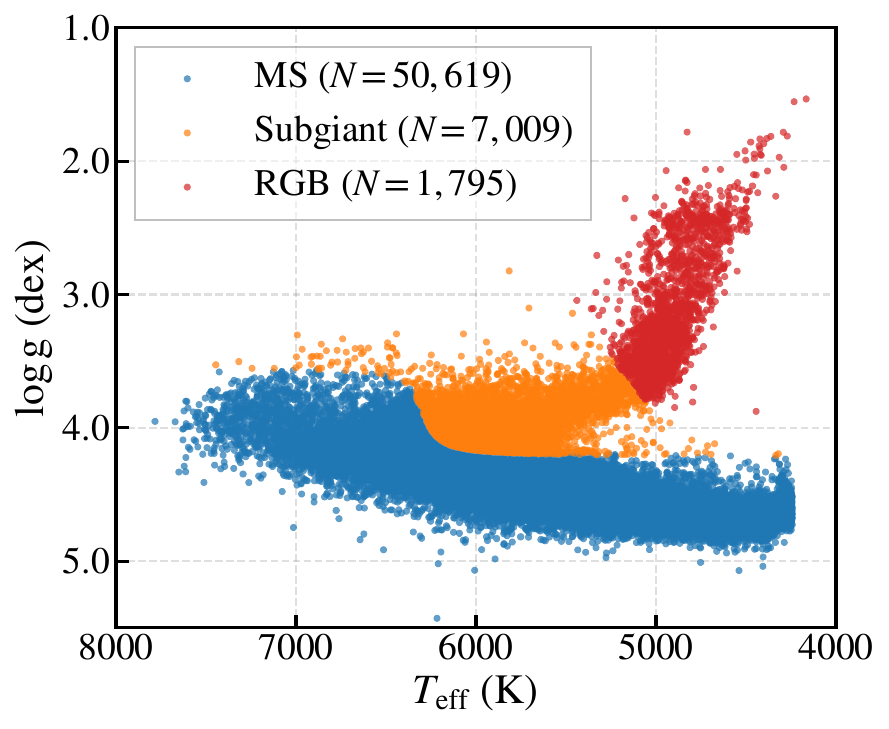}
\caption{Distribution of the wide--binary sample in the $\log g$–$T_{\mathrm{eff}}$ plane, classified according to the LAMOST atmospheric parameters of the primary component. Main-sequence (MS), subgiant, and red-giant-branch (RGB) systems are shown in blue, orange, and red, respectively; the corresponding numbers are given in the legend.}
\label{fig:evolstate}
\end{figure}

Although we initially retain MS--MS pairs, this classification is based on Gaia photometry and can include subgiants, giants, or other non-main-sequence contaminants \citep{ElBadry2021}. {To minimize evolutionary-stage-dependent variations in surface abundances,} we use the Python package \texttt{evolstate} \citep{Huber2016,Berger2018} to classify stellar evolutionary stages based on spectroscopic $\log g$ and $T_{\rm eff}$. Figure~\ref{fig:evolstate} shows classifications based only on primary-component parameters; including available secondary-component classifications yields the MS--MS sample of 53,329 systems.

Radial velocities are required to derive Galactic orbital properties, and provide an independent consistency check on binary membership. We adopt Gaia DR3 radial velocities when available and supplement the remaining $\sim 10\%$ of stars with measurements from the LAMOST DR9 low-resolution catalog. To account for the systematic offset between the LAMOST and Gaia radial-velocity scales \citep{Tian2015,Schonrich2017,Tian2019}, we apply a correction of $+5.38~{\rm km~s^{-1}}$ to the LAMOST values. For pairs with radial-velocity measurements for both components, we require the velocities to agree within $3\sigma_{\mathrm{rv}}$, where $\sigma_{\mathrm{rv}} = \sqrt{\sigma_{\mathrm{rv},1}^2 + \sigma_{\mathrm{rv},2}^2}$. This cut removes 278 systems, leaving 53,051 wide binaries. {The fraction of systems with \(\max(\mathrm{RUWE}_1,\mathrm{RUWE}_2)>1.4\) is higher among the RV-rejected pairs than among the RV-consistent pairs (60.1\% versus 43.4\%), consistent with unresolved higher-order multiplicity contributing to part of the RV-discrepant population \citep{ElBadry2021}.}

{Among the 53,051 systems retained after the RV-consistency cut, 2,179 have independent LAMOST spectra for both components. Their abundance differences show negligible median offsets and robust scatters of 0.078 dex in [Fe/H] and 0.039 dex in [$\alpha$/Fe] (see Appendix~\ref{app:component_consistency}: Figure~\ref{fig:binary_consistency} (b), (c)), supporting the use of mean abundances when both component measurements are available.}

\subsection{Subsample and population divisions}

To isolate the intrinsic metallicity dependence of the WBF from confounding factors, we stratify our sample by projected separation (s), effective temperature ($T_{\rm eff}$), and stellar population.

\subsubsection{Subsamples in \texorpdfstring{$s$}{s} and \texorpdfstring{$T_{\mathrm{eff}}$}{T\_eff}}

Because the WBF {varies with separation $s$ \citep{El-Badry2019,Tian2019a} and $T_{\rm eff}$ \citep{Niu2022}}, we divide the sample according to both parameters before measuring the WBF--[Fe/H] relation. For $s$, we adopt five separation intervals: \(0<s<200\,{\rm AU}\), \(200\,{\rm AU}<s<600\,{\rm AU}\), \(400\,{\rm AU}<s<800\,{\rm AU}\), \(600\,{\rm AU}<s<1000\,{\rm AU}\), and \(1000\,{\rm AU}<s<10000\,{\rm AU}\). The intermediate intervals are chosen with partial overlap so that the scale dependence of the WBF can be traced while retaining sufficient statistics in each bin. For $T_{\rm eff}$, we define three temperature regimes: $T_1$ ($4000\,{\rm K}<T_{\rm eff}<5400\,{\rm K}$), $T_2$ ($5400\,{\rm K}<T_{\rm eff}<6000\,{\rm K}$), and $T_3$ ($6000\,{\rm K}<T_{\rm eff}<7500\,{\rm K}$). This binning scheme enables the WBF--[Fe/H] relation to be measured within fixed ranges of both $s$ and $T_{\rm eff}$, while preserving adequate sample sizes.

\subsubsection{Low-\texorpdfstring{$\alpha$}{alpha} and high-\texorpdfstring{$\alpha$}{alpha} populations}

We further separate the sample by disk population because stars on the low- and high-$\alpha$ sequences have distinct chemical-enrichment histories, age distributions, and orbital properties \citep{Xiang2022}. In particular, chemically low-$\alpha$ stars on high-angular-momentum orbits are associated with the later, dynamically colder disk, whereas high-$\alpha$ stars predominantly trace the older, rapidly enriched disk population. Mixing these populations could therefore introduce population-dependent biases in the WBF--[Fe/H] relation, so we analyze the two populations separately.

To characterize the Galactic dynamical properties of the wide-binary sample, we combine the radial velocities described above with Gaia proper motions and parallaxes to calculate the angular momentum, $J_{\phi}$, and guiding radius, $R_{\rm g}$, using the \texttt{galpy} package \citep{Bovy2015} with the \texttt{MWPotential2014} Galactic potential. {The $R_{\rm g}$ is defined as the radius of the circular orbit with the same angular momentum as the star, satisfying $J_{\phi}=R_{\rm g}v_{\rm c}(R_{\rm g})$, where $v_{\rm c}(R_{\rm g})$ is the circular velocity in the adopted Galactic potential.} We adopt a solar Galactocentric distance of $R_{\odot}=8.178$ kpc \citep{GravityCollaboration2019}, a solar height above the plane of $Z_{\odot}=10$ pc \citep{Xiang2018} and a local-standard-of-rest (LSR) velocity of $220~{\rm km~s^{-1}}$. The solar motion with respect to the LSR is taken to be $(U_{\odot},V_{\odot},W_{\odot})=(9.58,10.52,7.01)~{\rm km~s^{-1}}$ \citep{Tian2015}. After excluding 94 systems without valid $J_{\phi}$ measurements, 52,957 wide binaries remain.

Following \citet{Xiang2022}, we define the low-$\alpha$ population using a joint cut in $J_{\phi}$, [Fe/H], and [$\alpha$/Fe]:

\begin{equation}
\begin{cases}
J_{\phi} > 1500\,\mathrm{kpc\,km\,s^{-1}}, \\[2pt]
[\alpha/\mathrm{Fe}] < 0.16,
[\mathrm{Fe}/\mathrm{H}] < -0.5, \\[2pt]
[\alpha/\mathrm{Fe}] < -0.16\,[\mathrm{Fe}/\mathrm{H}] + 0.08,
[\mathrm{Fe}/\mathrm{H}] \ge -0.5.
\end{cases}
\label{equ:population}
\end{equation}

\begin{figure}
\centering
\includegraphics[width=\linewidth]{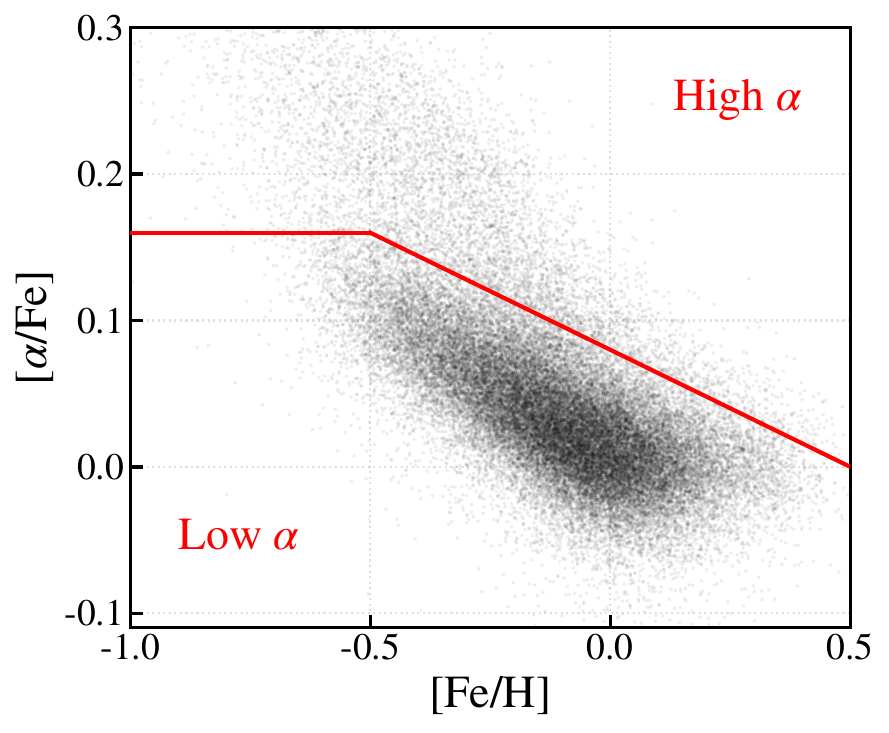}
\caption{Distribution of the low-$\alpha$ and high-$\alpha$ wide-binary populations in the [Fe/H]--[$\alpha$/Fe] plane. The red line marks the adopted chemical boundary between the two populations.}
\label{fig:alpha_plane}
\end{figure}

{This selection yields 42,746 low-$\alpha$ wide binaries. Of the remaining 10,211 systems, 6,202 lie above the chemical boundary and are classified as high-$\alpha$, while the other 4,009 are chemically low-$\alpha$ systems with $J_{\phi}\leq1500~{\rm kpc~km~s^{-1}}$ and are excluded (Figure~\ref{fig:alpha_plane}).}

\begin{table*}[htbp]
\centering
\caption{Summary of the Sample Selection and Population Construction \label{tab:sample_construction}}
\vspace{-6pt}
\small
\renewcommand{\arraystretch}{1.15}
\setlength{\tabcolsep}{6pt}

\begin{tabular}{c c c}
\hline\hline
Step & Selection criterion & Number \\
\hline

1 & \parbox[t]{0.7\textwidth}{Select MS--MS pairs with $R<0.1$ from the Gaia EDR3 wide-binary catalog of \citet{ElBadry2021}.} & 877,416 \\

2 & \parbox[t]{0.7\textwidth}{Crossmatch with the selected LAMOST DR9 DD-Payne catalog \citep{Zhang2025}.} & 61,983 \\

3 & \parbox[t]{0.7\textwidth}{Retain high-purity MS--MS wide binaries after evolutionary stage classification.} & 53,329 \\

4 & \parbox[t]{0.7\textwidth}{Compile available Gaia/LAMOST RVs, apply \(|\Delta RV|\le3\sigma_{RV}\) to systems with RVs for both components, and retain systems with valid $J_\phi$ measurements.} & 52,957 \\

5 & \parbox[t]{0.7\textwidth}{{Low-$\alpha$/High-$\alpha$/chemically low-$\alpha$, low-$J_{\phi}$ populations.}} & {42,746/6,202/4,009} \\

6 & \parbox[t]{0.7\textwidth}{Crossmatch with the LAMOST DR10 catalog \citep{Wang2025} to supplement age.} & {52,094} \\

\hline
\end{tabular}

\vspace{2mm}
\begin{minipage}{\textwidth}
\footnotesize
\textit{Note.} Steps 1--4 describe the sequential construction of the wide-binary sample. Step 5 divides this sample into different populations. Step 6 is an additional crossmatch used only for age-related analyses. The control sample is constructed independently from stars common to LAMOST and Gaia under the same basic quality cuts, and is not included in this table.
\end{minipage}
\end{table*}

\subsubsection{Age information for the wide-binary sample}

Stellar age is a key diagnostic for interpreting population-dependent trends in the WBF. However, reliable ages of dwarf stars are difficult to determine, because their atmospheric parameters evolve only weakly with age and isochrones of different ages are nearly degenerate, particularly for cool dwarfs. \citet{Wang2025} addressed this limitation by deriving spectroscopic ages for $\sim4$ million LAMOST DR10 dwarf stars using the \texttt{XGBoost} machine-learning algorithm. For high-signal-to-noise K-type dwarfs, they report typical age precisions of about 10--25\%, demonstrating the utility of the catalog for statistical studies of stellar ages. {We cross-match our wide-binary sample with this catalog and assign the available age to each system; when independent age estimates are available for both components, we adopt their mean. The \citet{Wang2025} catalog is limited to $4000\,{\rm K}<T_{\rm eff}<6800\,{\rm K}$ on its own temperature scale, so the age coverage is reduced toward the hot end of our $T_3$ ($6000\,{\rm K}<T_{\rm eff}<7500\,{\rm K}$). The effect of this mismatch is examined in Appendix~\ref{app:component_consistency}, panels (b) and (c) of Figure~\ref{fig:wbf_logg_age_robust}.}

{Age estimates are available for 52,094 wide-binary systems. Our subsequent age-based analyses focus on the low-$\alpha$ population, for which the median relative age uncertainties are 25.2\%, 15.6\%, and 16.8\% in $T_1$, $T_2$, and $T_3$, respectively. For 2,425 binaries with independent age measurements for both components, the component ages are centered close to the one-to-one relation, with a median difference of $-0.131$ Gyr and a robust scatter of 2.602 Gyr (Appendix~\ref{app:component_consistency}: Figure~\ref{fig:binary_consistency} (a)). Because the spectroscopic ages partly rely on abundance-sensitive information, we use them primarily to characterize population-level age trends rather than as precise, abundance-independent chronometers for individual stars.}

A summary of the sample-construction procedure is presented in Table~\ref{tab:sample_construction}.

\subsection{Measurement of the WBF}

The WBF has been quantified with different estimators in previous studies. \citet{Hwang2021} measured the absolute fraction, ${\rm WBF}=N_{\rm companion}/N_{\rm LAMOST}$, whereas \citet{El-Badry2019} and \citet{Niu2022} characterized its metallicity dependence by comparing wide binaries with a matched control sample. Following \citet{Niu2022}, we adopt the latter approach and refer to the resulting quantity as the relative WBF.

\begin{figure*}
\centering
\includegraphics[width=1\linewidth]{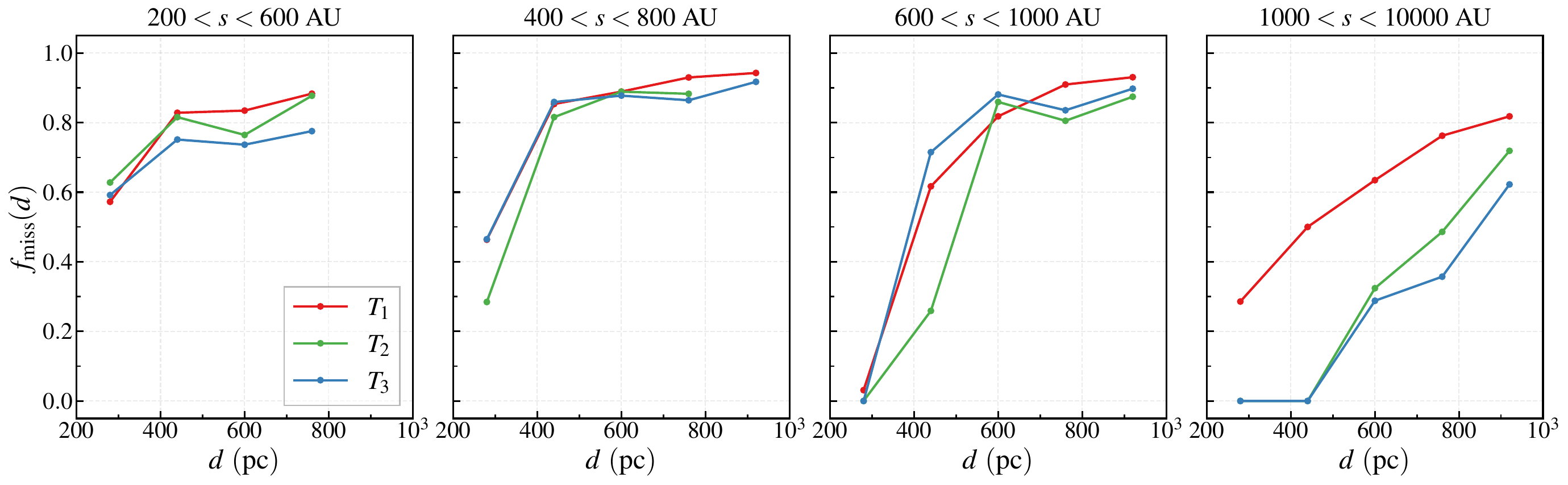}
\caption{Distance dependence of the missing fraction estimated using the twin-binary method, with binaries within 200 pc adopted as the reference sample. Different panels correspond to different $s$ intervals, and the red, green, and blue curves represent the $T_1$ ($4000\,{\rm K}<T_{\rm eff}<5400\,{\rm K}$), $T_2$ ($5400\,{\rm K}<T_{\rm eff}<6000\,{\rm K}$), and $T_3$ ($6000\,{\rm K}<T_{\rm eff}<7500\,{\rm K}$) subsamples, respectively. The missing fraction increases significantly with distance and shows broadly similar trends among three $T_{\rm eff}$ subsamples.}
\label{fig:completeness}
\end{figure*}

We construct the parent control sample from stars included in both the LAMOST DR9 DD-Payne catalog and Gaia EDR3, applying the same basic {spectroscopic, evolutionary-state, and population-selection} criteria as for the wide-binary sample. For each subsample defined by $s$, $T_{\rm eff}$, and stellar population, {we match the control sample to the wide-binary sample in the two-dimensional distance--$T_{\rm eff}$ plane using a $20\times20$ grid spanning the ranges of each subsample. Control stars are randomly selected without replacement at approximately a 1:1 ratio to the wide binaries; when a cell contains insufficient controls, stars from adjacent cells are used. The matched distance and $T_{\rm eff}$ distributions are verified using two-sample Kolmogorov--Smirnov tests.} The wide-binary and matched control samples are then divided into adaptive equal-number [Fe/H] bins over $-1 \leq {\rm [Fe/H]} \leq 0.5$, with the number of bins adjusted according to the size of the wide-binary subsample.

{For each retained [Fe/H] bin, $P_{\rm wb}$ and $P_{\rm ctrl}$ denote the fractions of the total wide-binary and matched control subsamples that fall within that bin, respectively.} The relative WBF is defined as
\begin{equation}
\boldsymbol{\mathcal{R}_{\rm WBF}}
=
\frac{P_{\rm wb}}{P_{\rm ctrl}}.
\label{equ:relative_wbf}
\end{equation}
{A value of $\boldsymbol{\mathcal{R}_{\rm WBF}}=1$ indicates equal normalized frequency in the wide-binary and control samples, while values above or below unity indicate relative enhancement or suppression, respectively. Uncertainties are obtained by combining 2000 bootstrap resamplings of the matched control sample, performed with replacement, with binomial uncertainties for the wide-binary sample; the same procedure is applied to the age-dependent analysis after restricting both samples to valid age measurements and rematching their distance and $T_{\rm eff}$ distributions.}

\subsection{Completeness Assessment}

The wide-binary sample may be affected by Gaia angular resolution and photometric limits, as well as by the target selection and quality cuts of LAMOST. These effects can preferentially remove systems with small angular separations or faint companions, particularly at large distances. To assess {the distance dependence of companion recovery}, we adopt the twin-binary method of \citet{Niu2022}.

Twin binaries are defined as systems with a Gaia $G$-band magnitude difference of $|\Delta G|<0.4$. {Because the two components have comparable brightness, such systems are less sensitive to the faint-companion limit. Under the assumption that the intrinsic mass-ratio distribution does not vary strongly at fixed primary mass and separation \citep{Moe2017,El-Badry2019}, an increase in the twin fraction with distance can indicate preferential loss of faint companions.}

The twin-binary fraction at distance $d$ is denoted by $f_{\rm twin}(d)$. The distance-dependent missing fraction is defined as
\begin{equation}
f_{\rm miss}(d)
=
1-\frac{f_{\rm twin}(d<200\,{\rm pc})}
        {f_{\rm twin}(d)},
\end{equation}
where $f_{\rm twin}(d<200\,{\rm pc})$ is the twin-binary fraction measured in the nearby reference sample. We compute $f_{\rm miss}$ separately for different $s$ and $T_{\rm eff}$ subsamples. 

Figure~\ref{fig:completeness} shows that the missing fraction generally increases with distance, as expected for incompleteness driven by faint companions. {However, similar distance dependence does not necessarily imply metallicity-independent companion recovery. We therefore additionally examine $\theta$ and $\Delta G$ as functions of [Fe/H] and repeat the relative-WBF--[Fe/H] analysis for a restricted sample. These tests are presented in Appendix~\ref{app:completeness} (Figure~\ref{fig:recovery_test}).}

\section{Results}
\label{results}

In this section, we present the metallicity dependence of the {relative WBF} across different $s$, $T_{\mathrm{eff}}$, and populations, revealing a temperature-dependent reversal in the low-$\alpha$ population but not in the high-$\alpha$ population. We further investigate how stellar age and $R_{\rm g}$ may contribute to this apparent reversal.

\subsection{Dependence of the {relative WBF} on \texorpdfstring{$s$}{s} and \texorpdfstring{$T_{\mathrm{eff}}$}{T\_eff} in the low-\texorpdfstring{$\alpha$}{alpha} and high-\texorpdfstring{$\alpha$}{alpha} populations}
\label{results-1}

\begin{figure*}[!htbp]
\centering
\includegraphics[width=\linewidth]{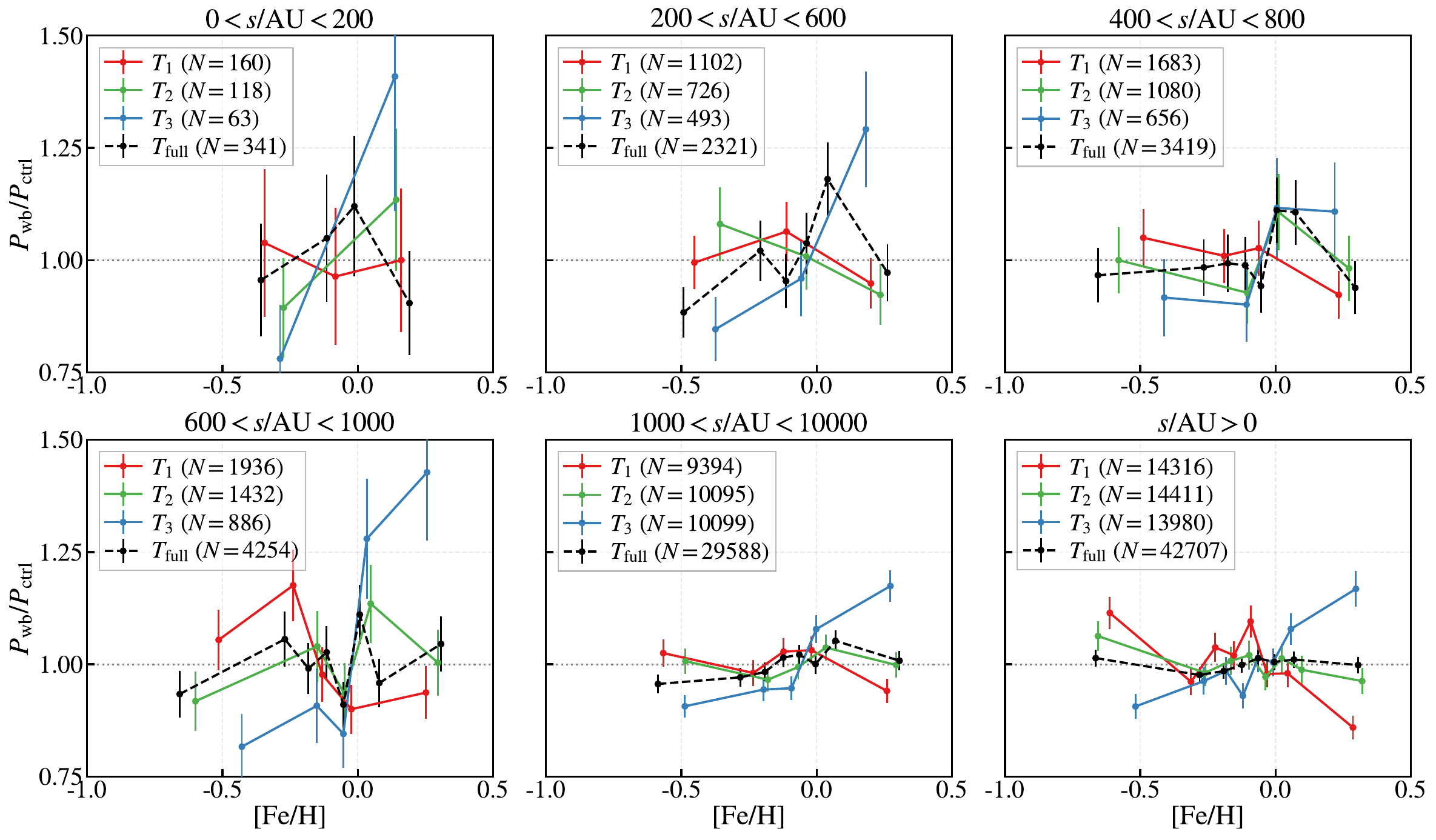}
\caption{{Relative WBF} as a function of [Fe/H] for the low-$\alpha$ population in different $s$ and $T_{\mathrm{eff}}$ subsamples. The red, green, and blue curves represent the $T_1$, $T_2$, and $T_3$ subsamples, respectively, while the black dashed curve shows the full-temperature sample over $4000$--$7500$ K. The numbers of wide binaries in each subsample are listed in the legend.}
\label{fig:lowalpha_wbf}
\end{figure*}

The low-$\alpha$ and high-$\alpha$ populations have distinct chemical enrichment histories and dynamical properties. Analyzing them separately reduces the mixing of stellar populations with different evolutionary origins, thereby enabling a more reliable characterization of the {relative WBF}--[Fe/H] relation. {Figure~\ref{fig:lowalpha_wbf} shows the relative-WBF--[Fe/H] relations for the low-$\alpha$ population across different $s$ and $T_{\rm eff}$ subsamples. The relation depends strongly on $T_{\rm eff}$.}

{For the $T_1$ subsample, the relative WBF generally decreases with increasing [Fe/H]. This negative metallicity dependence resembles the anti-correlation between binary fraction and [Fe/H] reported for close binaries \citep{Grether2007,Moe2019,Qiu2026} and at smaller separations by \citet{El-Badry2019}. Metallicity-dependent fragmentation and accretion during binary formation may imprint such a trend \citep{Machida2009,Bate2014,Bate2019}. In particular, the enhanced small-scale fragmentation found at low metallicity is most directly relevant to close companions, rather than to systems currently separated by thousands of AU. Some present-day wide binaries, however, may have originated in closer or higher-order multiple systems and subsequently reached wider orbits through dynamical interactions \citep{ElBadry2019Twins,Hwang2021,Hwang2022Twins,Shariat2025Triples}. This evolutionary pathway could preserve a formation-related metallicity imprint in part of the present-day $T_1$ wide-binary population. However, this metallicity dependence need not extend unchanged to very metal-poor populations. \citet{Shariat2025} found a wide-binary fraction beyond $5000\,{\rm AU}$ in Bo\"otes~I at $[\mathrm{Fe/H}]\simeq-2.5$ comparable to the Solar-neighborhood value at similar stellar masses and separations, whereas \citet{Lodieu2025} reported a lower companion fraction for metal-poor field stars over $8\lesssim s/{\rm AU}\lesssim10^4$. Differences in separation coverage, stellar-mass selection, formation environments, and observational completeness may contribute to these differing results. These comparisons therefore caution against extrapolating the negative $T_1$ relation measured over $-1 \leq {\rm [Fe/H]} \leq 0.5$ to substantially lower metallicities.}

\begin{figure*}[thbp]
\centering
\includegraphics[width=1\linewidth]{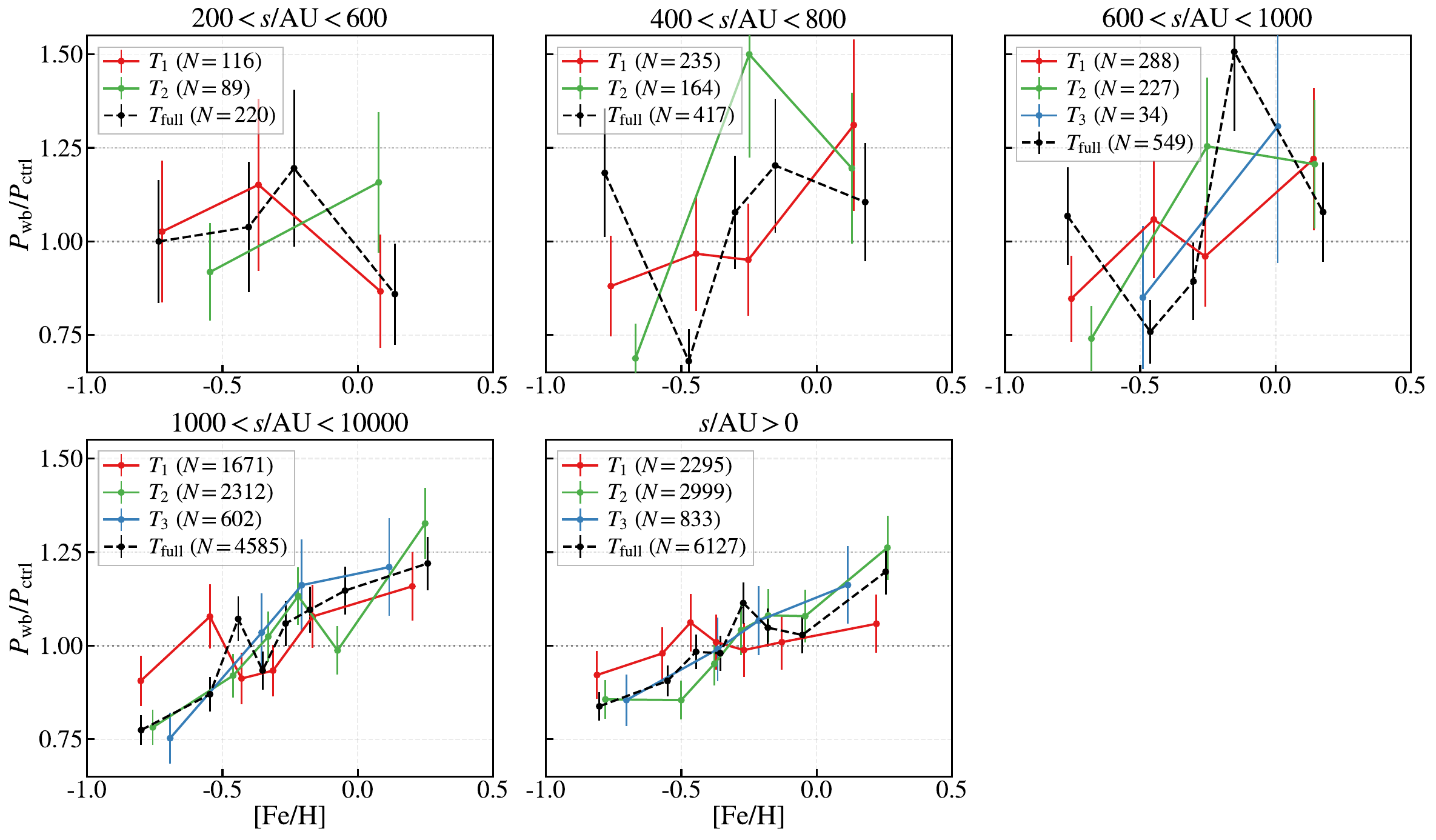}
\caption{{Relative WBF} as a function of [Fe/H] for the high-$\alpha$ population in different $s$ and $T_{\mathrm{eff}}$ subsamples. Colors and line styles are the same as in Figure~\ref{fig:lowalpha_wbf}. The $T_3$ subsample is not shown for $200\,{\rm AU}<s<600\,{\rm AU}$ and $400\,{\rm AU}<s<800\,{\rm AU}$ because of the small sample sizes ($N=15$ and $18$, respectively).}
\label{fig:highalpha_wbf}
\end{figure*}

As $T_{\rm eff}$ increases, the negative {relative-WBF}--[Fe/H] trend seen in $T_1$ becomes progressively weaker. The $T_2$ subsample shows an approximately flat relation, whereas the $T_3$ subsample exhibits a positive trend, with the {relative WBF} increasing toward higher [Fe/H]. {Because $T_1$ and $T_3$ exhibit trends of opposite sign, their superposition substantially weakens the metallicity dependence of $T_{\rm full}$ and produces mildly non-monotonic behavior in some separation intervals. The positive $T_3$ relation may reflect a combination of age, stellar-mass and birth-environment effects. Metal-rich $T_3$ stars may be younger and thus have experienced less cumulative dynamical disruption, potentially contributing to their higher relative WBF. At a given $T_{\rm eff}$, higher metallicity generally corresponds to a higher stellar mass for main-sequence stars at comparable evolutionary stages, owing to metallicity-dependent stellar opacity and structure \citep{Choi2016}. Since multiplicity increases with primary mass \citep{Duchne2013,Moe2017}, this coupling may contribute to the higher relative WBF of metal-rich $T_3$ stars. Birth environments may also play a role: wide-binary formation and early survival depend on cluster structure and dynamical evolution, with fewer wide systems observed in dense clusters than in loose associations \citep{Kouwenhoven2010,Deacon2020}. Metallicity-dependent differences in these environments could therefore contribute to the positive $T_3$ trend. The contrasting $T_1$ and $T_3$ relations establish a temperature-dependent reversal in the low-$\alpha$ population, consistent with \citet{Niu2022}.}

Figure~\ref{fig:highalpha_wbf} shows the {relative-WBF}--[Fe/H] relations of the high-$\alpha$ population in different $s$ and $T_{\mathrm{eff}}$ subsamples. At small to intermediate separations ($200\,{\rm AU}<s<1000\,{\rm AU}$), the {relative-WBF}--[Fe/H] relation of the high-$\alpha$ population displays relatively complex behavior. Because the number of high-$\alpha$ wide binaries in this $s$ range is limited, the statistical uncertainties of the individual $T_{\rm eff}$ subsamples are large, and the inferred trends are not robust. We therefore refrain from assigning a physical interpretation to this regime and focus instead on the larger-$s$ intervals, where the statistics are more constraining. At larger separation ($1000\,{\rm AU}<s<10000\,{\rm AU}$), as well as in the full-separation sample ($s>0$), {the relative WBF shows an overall increase with [Fe/H] across all three $T_{\rm eff}$ subsamples, despite some local fluctuations. In contrast to the low-$\alpha$ population, the high-$\alpha$ sample shows no systematic temperature-dependent reversal.}

{The high-$\alpha$ population is generally older and associated with the early phase of rapid star formation and chemical enrichment in the Galactic disk \citep{Haywood2013,Xiang2022,Xiang2024}. Within this population, lower [Fe/H] corresponds broadly to earlier stages of chemical evolution, when denser star-forming environments may have reduced the formation and early survival of wide binaries \citep{Hwang2021}. Consistent with this interpretation, wide binaries are less common in denser clusters than in looser associations of comparable age \citep{Deacon2020}. Furthermore, \citet{Hwang2022} reported similar wide-binary fractions between the thick disk and halo at fixed metallicity despite their different evolutionary histories, suggesting that the metallicity dependence is more closely associated with birth conditions than with disk-specific processes such as radial migration.}

Within comparable parameter ranges, our results are broadly consistent with previous studies. When restricted to the same $s$ and $T_{\mathrm{eff}}$ ranges as \citet{Hwang2021}, the {relative-WBF}--[Fe/H] relations of our low-$\alpha$ and high-$\alpha$ populations are consistent with their thin- and thick-disk samples, respectively (see Appendix~\ref{app:hwang2021}: Figure~\ref{fig:compare_hwang}.) {Together with the separation dependence reported by \citet{El-Badry2019} and the temperature dependence found by \citet{Niu2022}, these results indicate that the observed relative-WBF--[Fe/H] relation varies with separation, $T_{\rm eff}$, and stellar population.}

\begin{figure*}
\centering
\includegraphics[width=1\linewidth]{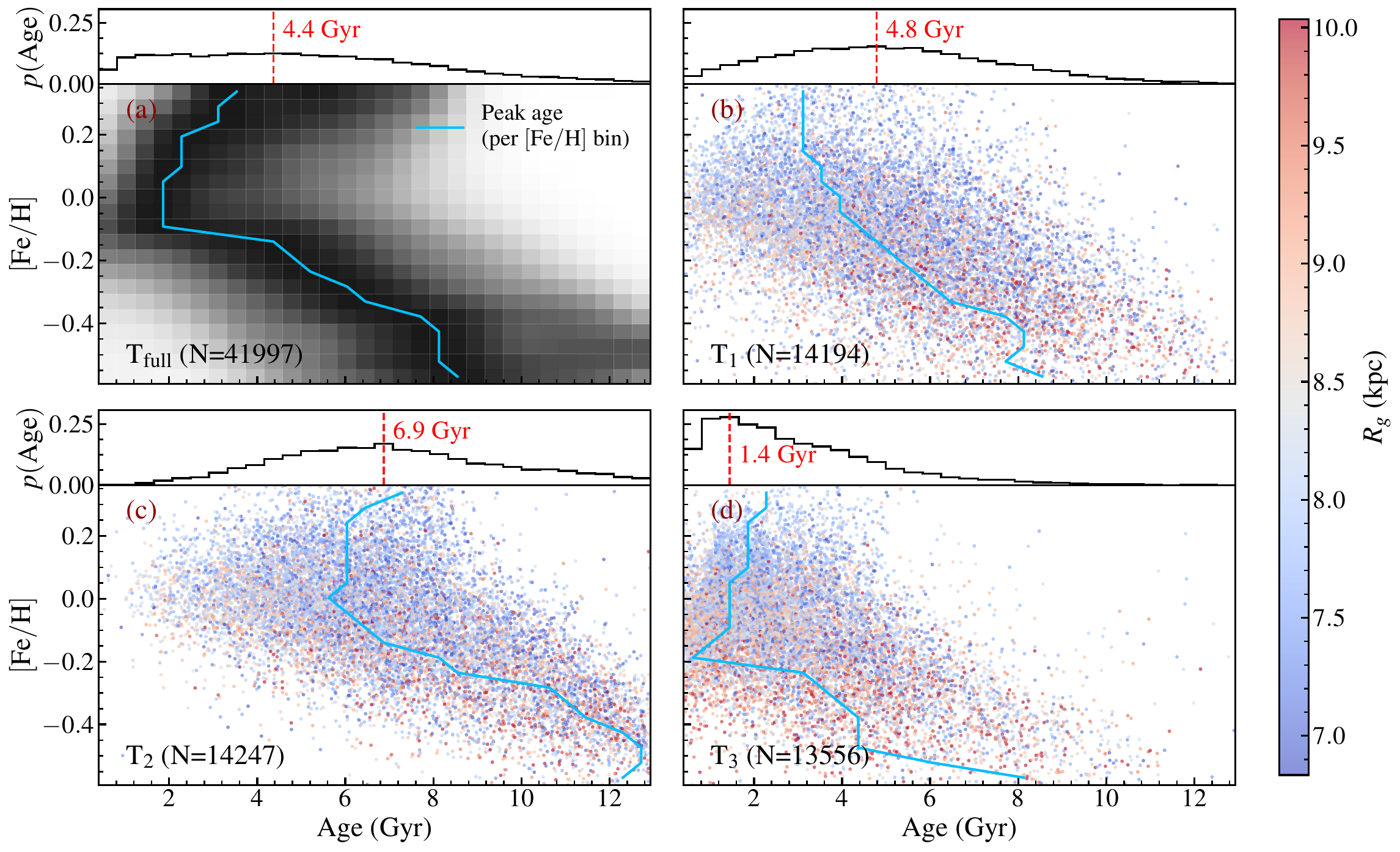}
\caption{Distribution of the low-$\alpha$ wide-binary sample with $s>0$ AU in the age--[Fe/H] plane. Panel (a) shows the density distribution of the full-temperature sample, $T_{\rm full}$, while panels (b)--(d) show the distributions of the $T_1$, $T_2$, and $T_3$ subsamples, respectively, with points color-coded by $R_{\rm g}$. The blue curve in each panel traces the peak age in each [Fe/H] bin. The histogram above each panel shows the normalized age distribution, $p({\rm Age})$, and the red dashed line marks its peak.}
\label{fig:age_feh_lowalpha}
\end{figure*}

\subsection{Disentangling the {relative-WBF}–[Fe/H] reversal in the low-\texorpdfstring{$\alpha$}{alpha} population}
\label{sec:age_feh}

To further understand the temperature-dependent reversal in the low-$\alpha$ population, we examine the joint distribution of wide binaries in age, [Fe/H], and $R_{\rm g}$. 

Figure~\ref{fig:age_feh_lowalpha} shows the distributions of different $T_{\mathrm{eff}}$ subsamples in the age--[Fe/H] plane. The $T_{\rm full}$ sample in Figure~\ref{fig:age_feh_lowalpha}(a) exhibits a pronounced ‘V-shaped’ age--[Fe/H] structure. The typical age reaches a minimum near ${\rm [Fe/H]}\sim-0.1$. Stars become progressively older toward both the metal-poor and metal-rich sides of this minimum. This non-monotonic behavior is consistent with previous studies of the age--metallicity relation of the Galactic disk \citep{Feuillet2018, Feuillet2019, Lu2022, Xiang2022, Sun2023}. Figures~\ref{fig:age_feh_lowalpha}(b)--(d) further show that {the two metallicity branches also differ in $R_{\rm g}$: metal-poor systems preferentially occupy larger $R_g$, whereas metal-rich systems tend to have smaller $R_g$. This radial structure is consistent with the effects of radial mixing on the local age--metallicity distribution \citep{Feuillet2018,Feuillet2019}.}

\begin{figure}[ht]
\centering
\includegraphics[width=1\linewidth]{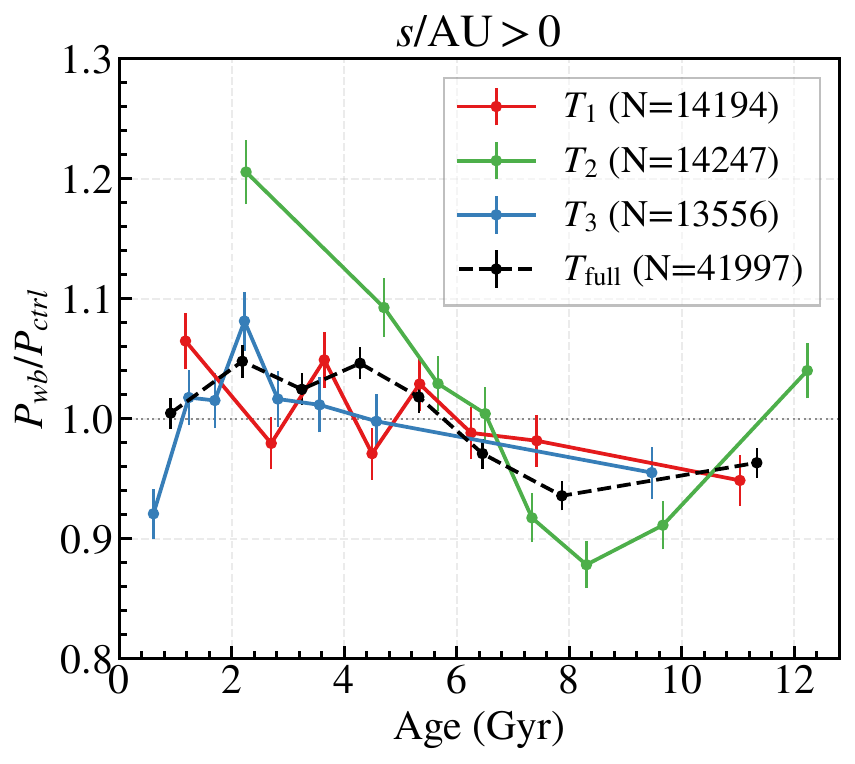}
\caption{{Relative WBF} as a function of age for the low-$\alpha$ population with $s>0\,{\rm AU}$. The red, green, and blue solid curves represent the $T_1$, $T_2$, and $T_3$ subsamples, respectively, while the black dashed curve denotes the full-temperature sample, $T_{\rm full}$.}
\label{fig:wbf_age_lowalpha}
\end{figure}

The different $T_{\rm eff}$ subsamples occupy distinct regions of the age--[Fe/H] plane. The peak ages are {$4.8~{\rm Gyr}$ for $T_1$}, approximately {$6.9~{\rm Gyr}$ for $T_2$}, and approximately $1.4~{\rm Gyr}$ for $T_3$. {The $T_3$ sample is concentrated toward younger ages at higher [Fe/H], whereas $T_1$ spans a broader age range. Given the distinct age--[Fe/H] distributions of the three temperature subsamples, we examine their relative WBF as a function of age. Figure~\ref{fig:wbf_age_lowalpha} shows that the relative WBF decreases with age in all three subsamples over 0--14 Gyr, without the temperature-dependent reversal observed in their relative-WBF--[Fe/H] relations.}

\begin{figure*}
\centering
\includegraphics[width=1\linewidth]{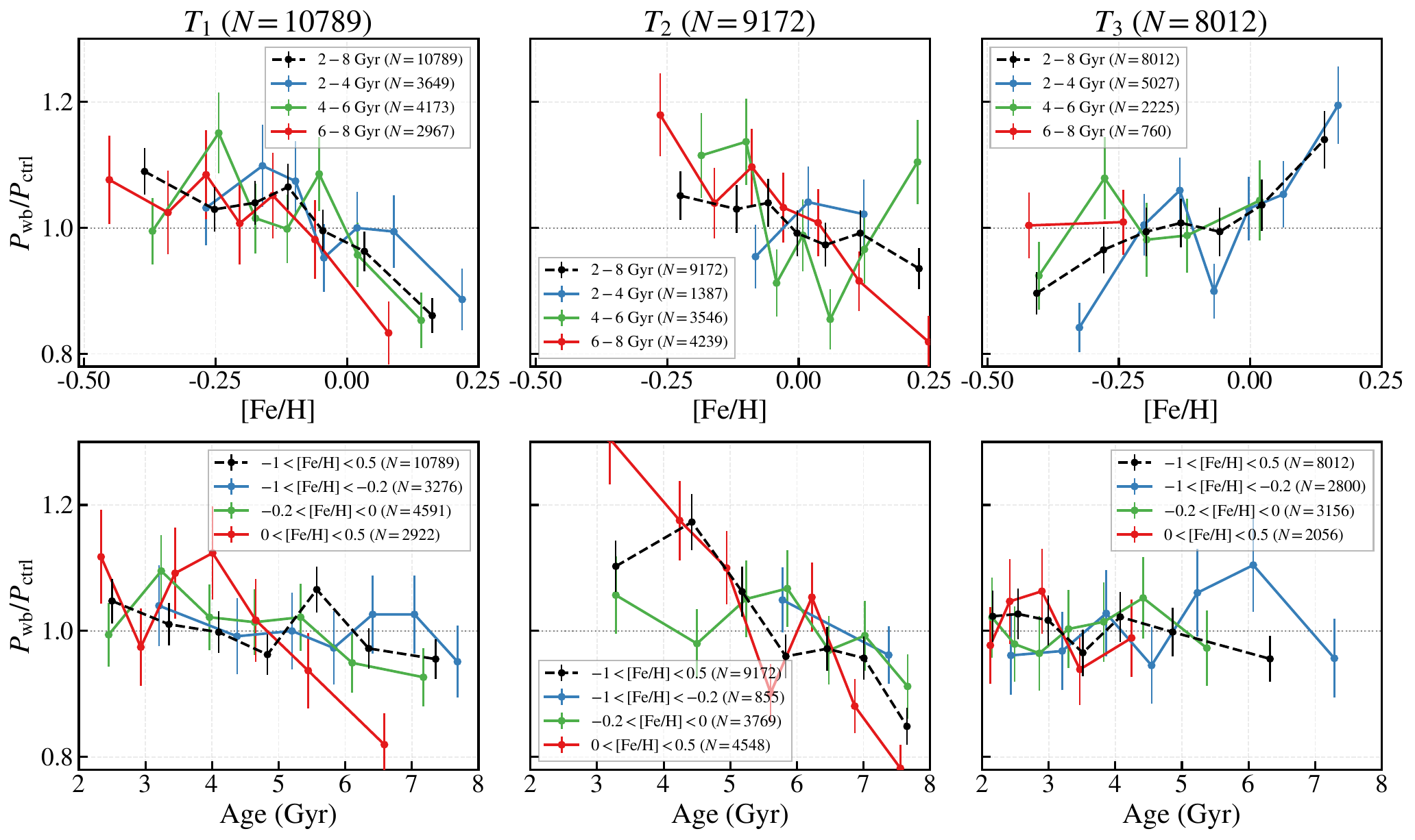}
\caption{{Relative WBF versus [Fe/H] within common age intervals (top) and versus age within [Fe/H] intervals (bottom) for low-$\alpha$ systems with $s>0\,{\rm AU}$ and $2\leq {\rm Age}\leq8$ Gyr. Columns show $T_1$, $T_2$, and $T_3$; black dashed curves refer to the corresponding full 2--8 Gyr samples. Colors denote the age or [Fe/H] intervals indicated in each panel.}}
\label{fig:joint_age_feh}
\end{figure*}

{To determine whether differences in age distribution can account for the metallicity reversal, we further compare the relative-WBF--[Fe/H] relations within common age intervals. Figure~\ref{fig:joint_age_feh} shows the relative WBF within common age and [Fe/H] intervals. In the upper panels, the relative WBF generally decreases with [Fe/H] for $T_1$ but increases for $T_3$ within comparable age intervals, whereas $T_2$ shows a weaker dependence. The $T_3$ trend is less well constrained at 6--8 Gyr because of the smaller sample size. In the lower panels, the relative WBF generally declines with age for several $T_1$ and $T_2$ subsamples within individual [Fe/H] intervals, while $T_3$ exhibits a weaker and less systematic dependence. The persistence of the opposing metallicity trends of $T_1$ and $T_3$ within the same age bins indicates that differences in their age--[Fe/H] distributions cannot fully account for the temperature-dependent reversal.}

{The overall decline in relative WBF with age could reflect the cumulative disruption of wide binaries by stellar encounters, molecular clouds, and the Galactic tidal field \citep{Heggie1975,Weinberg1987,Jiang2010}. These perturbations may alter orbital eccentricities and orientations without unbinding the systems, whereas a reduction in the intrinsic fraction of bound binaries requires disruption. If cumulative disruption dominates the observed age trend, the decline should strengthen toward larger separations, where binaries are more weakly bound. However, our separation-resolved analysis shows no systematic strengthening of the decline from $200\,{\rm AU}<s<1000\,{\rm AU}$ to $1000\,{\rm AU}<s<10000\,{\rm AU}$ (Appendix~\ref{app:separation_age}; Figure~\ref{fig:wbf_age_sep}), consistent with \citet{Hwang2021}. The absence of a systematically stronger decline at larger separations suggests that cumulative field disruption may not be the sole driver of the observed age dependence.}

{These results provide further constraints on the physical interpretation of the temperature-dependent reversal. Although the younger ages of metal-rich $T_3$ systems may favor their dynamical survival, the persistence of the positive metallicity trend within comparable age intervals indicates that age-related survival cannot fully account for this relation. The negative $T_1$ relation may retain a formation-related metallicity imprint through the dynamical widening of initially closer or higher-order systems, whereas the positive $T_3$ relation may additionally reflect mass-dependent multiplicity and differences in birth environments, as discussed in Section~\ref{results-1}. The reversal may therefore arise from different contributions of binary formation and subsequent evolution across the temperature subsamples.}

\section{Summary}
\label{summary}

We combine the Gaia EDR3 wide-binary catalog with LAMOST stellar parameters to construct a sample of 52,957 high-purity main-sequence wide binaries within 1 kpc. Using distance- and $T_{\rm eff}$-matched control samples, we measure the {relative WBF} as a function of [Fe/H], $s$, and $T_{\rm eff}$ for the low-$\alpha$ and high-$\alpha$ populations.

For the low-$\alpha$ population at $s>200$ AU, the {relative-WBF}--[Fe/H] relation reverses with $T_{\rm eff}$: it is negative for cool stars ($4000\,{\rm K}<T_{\rm eff}<5400\,{\rm K}$), approximately flat at intermediate temperatures, and positive for hot stars ($6000\,{\rm K}<T_{\rm eff}<7500\,{\rm K}$). No analogous reversal is found in the high-$\alpha$ population, whose {relative WBF} generally increases with [Fe/H] for $1000\,{\rm AU}<s<10000\,{\rm AU}$.

The low-$\alpha$ population exhibits a V-shaped age--metallicity distribution that covaries with $R_{\rm g}$, while the three temperature subsamples have markedly different age distributions. {The relative WBF declines with age across all three temperature subsamples, yet the opposing metallicity trends of $T_1$ and $T_3$ persist within comparable age intervals, indicating that differences in age distribution cannot fully explain the reversal. The age-related decline shows no systematic strengthening with increasing separation, suggesting that cumulative field disruption is not its sole driver. The negative $T_1$ relation may retain a formation-related metallicity imprint, whereas the positive $T_3$ relation may reflect metallicity-dependent stellar-mass selection and differences in birth environments.} More precise stellar ages, improved orbital constraints, and samples extending beyond the solar neighborhood will be required to fully quantify the relative contributions of formation physics and dynamical survival---a task that will be further empowered by future Gaia releases and large-scale spectroscopic surveys.

\begin{acknowledgements}
This work was supported by the National Natural Science Foundation of China (Grants No.12403037, 12373033) and the Key Project of Zhejiang Provincial Natural Science Foundation (No. ZCLZ25A0301). This work is based on data acquired through the Guoshoujing Telescope. Guoshoujing Telescope (the Large Sky Area Multi-Object Fiber Spectroscopic Telescope; LAMOST) is a National Major Scientific Project built by the Chinese Academy of Sciences. Funding for the project has been provided by the National Development and Reform Commission. LAMOST is operated and managed by the National Astronomical Observatories, Chinese Academy of Sciences. This work has made use of data from the European Space Agency (ESA) mission Gaia (\url{https://www.cosmos.esa.int/gaia}), processed by the Gaia Data Processing and Analysis Consortium (DPAC, \url{https://www.cosmos.esa.int/web/gaia/dpac/consortium}). Funding for the DPAC has been provided by national institutions, in particular the institutions participating in the Gaia Multilateral Agreement. 
\end{acknowledgements}

% Bibliography
\bibliographystyle{aasjournalv7}
\bibliography{Biblio}

\begin{thebibliography}{}
\expandafter\ifx\csname natexlab\endcsname\relax\def\natexlab#1{#1}\fi
\providecommand{\url}[1]{\href{#1}{#1}}
\providecommand{\dodoi}[1]{doi:~\href{http://doi.org/#1}{\nolinkurl{#1}}}
\providecommand{\doeprint}[1]{\href{http://ascl.net/#1}{\nolinkurl{http://ascl.net/#1}}}
\providecommand{\doarXiv}[1]{\href{https://arxiv.org/abs/#1}{\nolinkurl{https://arxiv.org/abs/#1}}}

% type= article
\bibitem[{R. Abuter {et~al.}(2019)Abuter, Amorim, Baub\"{o}ck, Berger, Bonnet,
  Brandner, Clénet, Coudé~du Foresto, de~Zeeuw, Dexter, Duvert, Eckart,
  Eisenhauer, F\"{o}rster~Schreiber, Garcia, Gao, Gendron, Genzel, Gerhard,
  Gillessen, Habibi, Haubois, Henning, Hippler, Horrobin, Jiménez-Rosales,
  Jocou, Kervella, Lacour, Lapeyrère, Le~Bouquin, Léna, Ott, Paumard,
  Perraut, Perrin, Pfuhl, Rabien, Rodriguez~Coira, Rousset, Scheithauer,
  Sternberg, Straub, Straubmeier, Sturm, Tacconi, Vincent, von Fellenberg,
  Waisberg, Widmann, Wieprecht, Wiezorrek, Woillez, \&
  Yazici}]{GravityCollaboration2019}
Abuter, R., Amorim, A., Baub\"{o}ck, M., {et~al.} 2019, \bibinfo{title}{A
  geometric distance measurement to the Galactic center black hole with 0.3%
  uncertainty,} \aap, 625, L10, \dodoi{10.1051/0004-6361/201935656}

% type= article
\bibitem[{J.~J. Andrews {et~al.}(2019)Andrews, Anguiano, Chanamé, Ag\"{u}eros,
  Lewis, Hayes, \& Majewski}]{Andrews2019}
Andrews, J.~J., Anguiano, B., Chanamé, J., {et~al.} 2019,
  \bibinfo{title}{Using APOGEE Wide Binaries to Test Chemical Tagging with
  Dwarf Stars,} \apj, 871, 42, \dodoi{10.3847/1538-4357/aaf502}

% type= article
\bibitem[{J.~J. Andrews {et~al.}(2017)Andrews, Chanamé, \&
  Ag\"{u}eros}]{Andrews2017}
Andrews, J.~J., Chanamé, J., \& Ag\"{u}eros, M.~A. 2017, \bibinfo{title}{Wide
  binaries in Tycho-Gaia II: metallicities, abundances and prospects for
  chemical tagging,} \mnras, 473, 5393–5406, \dodoi{10.1093/mnras/stx2685}

% type= article
\bibitem[{M.~R. Bate(2014)Bate}]{Bate2014}
Bate, M.~R. 2014, \bibinfo{title}{The statistical properties of stars and their
  dependence on metallicity: the effects of opacity,} \mnras, 442, 285–313,
  \dodoi{10.1093/mnras/stu795}

% type= article
\bibitem[{M.~R. Bate(2019)Bate}]{Bate2019}
Bate, M.~R. 2019, \bibinfo{title}{The statistical properties of stars and their
  dependence on metallicity,} \mnras, 484, 2341–2361,
  \dodoi{10.1093/mnras/stz103}

% type= article
\bibitem[{T.~A. Berger {et~al.}(2018)Berger, Huber, Gaidos, \& van
  Saders}]{Berger2018}
Berger, T.~A., Huber, D., Gaidos, E., \& van Saders, J.~L. 2018,
  \bibinfo{title}{Revised Radii of Kepler Stars and Planets Using Gaia Data
  Release 2,} \apj, 866, 99, \dodoi{10.3847/1538-4357/aada83}

% type= article
\bibitem[{J. {Bovy}(2015){Bovy}}]{Bovy2015}
{Bovy}, J. 2015, \bibinfo{title}{{galpy: A Python Library for Galactic
  Dynamics},} \apjs, 216, 29, \dodoi{10.1088/0067-0049/216/2/29}

% type= article
\bibitem[{A.~G.~A. Brown {et~al.}(2021)Brown, Vallenari, Prusti, de~Bruijne,
  Babusiaux, Biermann, Creevey, Evans, Eyer, Hutton, Jansen, Jordi, Klioner,
  Lammers, Lindegren, Luri, Mignard, Panem, Pourbaix, Randich, Sartoretti,
  Soubiran, Walton, Arenou, Bailer-Jones, Bastian, Cropper, Drimmel, Katz,
  Lattanzi, van Leeuwen, Bakker, Cacciari, Castañeda, De~Angeli, Ducourant,
  Fabricius, Fouesneau, Frémat, Guerra, Guerrier, Guiraud,
  Jean-Antoine~Piccolo, Masana, Messineo, Mowlavi, Nicolas, Nienartowicz,
  Pailler, Panuzzo, Riclet, Roux, Seabroke, Sordo, Tanga, Thévenin,
  Gracia-Abril, Portell, Teyssier, Altmann, Andrae, Bellas-Velidis, Benson,
  Berthier, Blomme, Brugaletta, Burgess, Busso, Carry, Cellino, Cheek,
  Clementini, Damerdji, Davidson, Delchambre, Dell’Oro,
  Fernández-Hernández, Galluccio, García-Lario, Garcia-Reinaldos,
  González-Núñez, Gosset, Haigron, Halbwachs, Hambly, Harrison,
  Hatzidimitriou, Heiter, Hernández, Hestroffer, Hodgkin, Holl, Janßen,
  Jevardat~de Fombelle, Jordan, Krone-Martins, Lanzafame, L\"{o}ffler, Lorca,
  Manteiga, Marchal, Marrese, Moitinho, Mora, Muinonen, Osborne, Pancino,
  Pauwels, Petit, Recio-Blanco, Richards, Riello, Rimoldini, Robin, Roegiers,
  Rybizki, Sarro, Siopis, Smith, Sozzetti, Ulla, Utrilla, van Leeuwen, van
  Reeven, Abbas, Abreu~Aramburu, Accart, Aerts, Aguado, Ajaj, Altavilla,
  Álvarez, Álvarez Cid-Fuentes, Alves, Anderson, Anglada~Varela, Antoja,
  Audard, Baines, Baker, Balaguer-Núñez, Balbinot, Balog, Barache, Barbato,
  Barros, Barstow, Bartolomé, Bassilana, Bauchet, Baudesson-Stella, Becciani,
  Bellazzini, Bernet, Bertone, Bianchi, Blanco-Cuaresma, Boch, Bombrun,
  Bossini, Bouquillon, Bragaglia, Bramante, Breedt, Bressan, Brouillet,
  Bucciarelli, Burlacu, Busonero, Butkevich, Buzzi, Caffau, Cancelliere,
  Cánovas, Cantat-Gaudin, Carballo, Carlucci, Carnerero, Carrasco,
  Casamiquela, Castellani, Castro-Ginard, Castro~Sampol, Chaoul, Charlot,
  Chemin, Chiavassa, Cioni, Comoretto, Cooper, Cornez, Cowell, Crifo, Crosta,
  Crowley, Dafonte, Dapergolas, David, David, de~Laverny, De~Luise, De~March,
  De~Ridder, de~Souza, de~Teodoro, de~Torres, del Peloso, del Pozo, Delbo,
  Delgado, Delgado, Delisle, Di~Matteo, Diakite, Diener, Distefano, Dolding,
  Eappachen, Edvardsson, Enke, Esquej, Fabre, Fabrizio, Faigler, Fedorets,
  Fernique, Fienga, Figueras, Fouron, Fragkoudi, Fraile, Franke, Gai, Garabato,
  Garcia-Gutierrez, García-Torres, Garofalo, Gavras, Gerlach, Geyer, Giacobbe,
  Gilmore, Girona, Giuffrida, Gomel, Gomez, Gonzalez-Santamaria,
  González-Vidal, Granvik, Gutiérrez-Sánchez, Guy, Hauser, Haywood, Helmi,
  Hidalgo, Hilger, Hładczuk, Hobbs, Holland, Huckle, Jasniewicz, Jonker,
  Juaristi~Campillo, Julbe, Karbevska, Kervella, Khanna, Kochoska, Kontizas,
  Kordopatis, Korn, Kostrzewa-Rutkowska, Kruszyńska, Lambert, Lanza, Lasne,
  Le~Campion, Le~Fustec, Lebreton, Lebzelter, Leccia, Leclerc, Lecoeur-Taibi,
  Liao, Licata, Lindstrøm, Lister, Livanou, Lobel, Madrero~Pardo, Managau,
  Mann, Marchant, Marconi, Marcos~Santos, Marinoni, Marocco, Marshall,
  Martin~Polo, Martín-Fleitas, Masip, Massari, Mastrobuono-Battisti, Mazeh,
  McMillan, Messina, Michalik, Millar, Mints, Molina, Molinaro, Molnár,
  Montegriffo, Mor, Morbidelli, Morel, Morris, Mulone, Munoz, Muraveva, Murphy,
  Musella, Noval, Ordénovic, Orrù, Osinde, Pagani, Pagano, Palaversa,
  Palicio, Panahi, Pawlak, Peñalosa~Esteller, Penttil\"{a}, Piersimoni,
  Pineau, Plachy, Plum, Poggio, Poretti, Poujoulet, Prša, Pulone, Racero,
  Ragaini, Rainer, Raiteri, Rambaux, Ramos, Ramos-Lerate, Re~Fiorentin, Regibo,
  Reylé, Ripepi, Riva, Rixon, Robichon, Robin, Roelens, Rohrbasser,
  Romero-Gómez, Rowell, Royer, Rybicki, Sadowski, Sagristà~Sellés, Sahlmann,
  Salgado, Salguero, Samaras, Sanchez~Gimenez, Sanna, Santoveña, Sarasso,
  Schultheis, Sciacca, Segol, Segovia, Ségransan, Semeux, Shahaf, Siddiqui,
  Siebert, Siltala, Slezak, Smart, Solano, Solitro, Souami, Souchay, Spagna,
  Spoto, Steele, Steidelm\"{u}ller, Stephenson, S\"{u}veges, Szabados,
  Szegedi-Elek, Taris, Tauran, Taylor, Teixeira, Thuillot, Tonello, Torra,
  Torra, Turon, Unger, Vaillant, van Dillen, Vanel, Vecchiato, Viala, Vicente,
  Voutsinas, Weiler, Wevers, Wyrzykowski, Yoldas, Yvard, Zhao, Zorec, Zucker,
  Zurbach, \& Zwitter}]{Gaia2021}
Brown, A. G.~A., Vallenari, A., Prusti, T., {et~al.} 2021, \bibinfo{title}{Gaia
  Early Data Release 3: Summary of the contents and survey properties,} \aap,
  649, A1, \dodoi{10.1051/0004-6361/202039657}

% type= misc
\bibitem[{A. Caputo {et~al.}(2026)Caputo, Tomaselli, \& Hamilton}]{Caputo2026}
Caputo, A., Tomaselli, G.~M., \& Hamilton, C. 2026, Evolution of Binaries Under
  Stochastic Perturbations, arXiv, \dodoi{10.48550/ARXIV.2607.12011}

% type= article
\bibitem[{J. Chanamé \& I. Ramírez(2012)Chanamé \& Ramírez}]{Chanam2012}
Chanamé, J., \& Ramírez, I. 2012, \bibinfo{title}{TOWARD PRECISE AGES FOR
  SINGLE STARS IN THE FIELD. GYROCHRONOLOGY CONSTRAINTS AT SEVERAL Gyr USING
  WIDE BINARIES. I. AGES FOR INITIAL SAMPLE,} \apj, 746, 102,
  \dodoi{10.1088/0004-637x/746/1/102}

% type= article
\bibitem[{X. Chen {et~al.}(2024)Chen, Liu, \& Han}]{Chen2024}
Chen, X., Liu, Z., \& Han, Z. 2024, \bibinfo{title}{Binary stars in the new
  millennium,} Progress in Particle and Nuclear Physics, 134, 104083,
  \dodoi{10.1016/j.ppnp.2023.104083}

% type= article
\bibitem[{J. Choi {et~al.}(2016)Choi, Dotter, Conroy, Cantiello, Paxton, \&
  Johnson}]{Choi2016}
Choi, J., Dotter, A., Conroy, C., {et~al.} 2016, \bibinfo{title}{MESA
  ISOCHRONES AND STELLAR TRACKS (MIST). I. SOLAR-SCALED MODELS,} \apj, 823,
  102, \dodoi{10.3847/0004-637x/823/2/102}

% type= article
\bibitem[{J.~A. Correa-Otto \& R.~A. Gil-Hutton(2017)Correa-Otto \&
  Gil-Hutton}]{CorreaOtto2017}
Correa-Otto, J.~A., \& Gil-Hutton, R.~A. 2017, \bibinfo{title}{Galactic
  perturbations on the population of wide binary stars with exoplanets,} \aap,
  608, A116, \dodoi{10.1051/0004-6361/201731229}

% type= article
\bibitem[{X.-Q. Cui {et~al.}(2012)Cui, Zhao, Chu, Li, Li, Zhang, Su, Yao, Wang,
  Xing, Li, Zhu, Wang, Gu, Luo, Xu, Zhang, Liu, Zhang, Yang, Cao, Chen, Chen,
  Chen, Chen, Chu, Feng, Gong, Hou, Hu, Hu, Hu, Jia, Jiang, Jiang, Jiang, Jin,
  Li, Li, Li, Liu, Liu, Lu, Mao, Men, Qi, Qi, Shi, Tang, Tao, Wang, Wang, Wang,
  Wang, Wang, Wang, Wang, Wang, Wang, Wang, Wang, Wang, Xu, Xu, Yang, Yu, Yuan,
  Yuan, Zhai, Zhang, Zhang, Zhang, Zhao, Zhou, Zhou, Zhu, \& Zou}]{Cui2012}
Cui, X.-Q., Zhao, Y.-H., Chu, Y.-Q., {et~al.} 2012, \bibinfo{title}{The Large
  Sky Area Multi-Object Fiber Spectroscopic Telescope (LAMOST),} RAA, 12,
  1197–1242, \dodoi{10.1088/1674-4527/12/9/003}

% type= article
\bibitem[{N.~R. Deacon \& A.~L. Kraus(2020)Deacon \& Kraus}]{Deacon2020}
Deacon, N.~R., \& Kraus, A.~L. 2020, \bibinfo{title}{Wide binaries are rare in
  open clusters,} \mnras, 496, 5176–5200, \dodoi{10.1093/mnras/staa1877}

% type= article
\bibitem[{G. Duch\^ene \& A. Kraus(2013)Duch\^ene \& Kraus}]{Duchne2013}
Duch\^ene, G., \& Kraus, A. 2013, \bibinfo{title}{Stellar Multiplicity,} \araa,
  51, 269–310, \dodoi{10.1146/annurev-astro-081710-102602}

% type= article
\bibitem[{K. El-Badry(2024)El-Badry}]{ElBadry2024}
El-Badry, K. 2024, \bibinfo{title}{Gaia’s binary star renaissance,} \nar, 98,
  101694, \dodoi{10.1016/j.newar.2024.101694}

% type= article
\bibitem[{K. {El-Badry} \& H.-W. {Rix}(2019){El-Badry} \& {Rix}}]{El-Badry2019}
{El-Badry}, K., \& {Rix}, H.-W. 2019, \bibinfo{title}{{The wide binary fraction
  of solar-type stars: emergence of metallicity dependence at a < 200 au},}
  \mnras, 482, L139, \dodoi{10.1093/mnrasl/sly206}

% type= article
\bibitem[{K. {El-Badry} {et~al.}(2021){El-Badry}, {Rix}, \&
  {Heintz}}]{ElBadry2021}
{El-Badry}, K., {Rix}, H.-W., \& {Heintz}, T.~M. 2021, \bibinfo{title}{{A
  million binaries from Gaia eDR3: sample selection and validation of Gaia
  parallax uncertainties},} \mnras, 506, 2269, \dodoi{10.1093/mnras/stab323}

% type= article
\bibitem[{K. El-Badry {et~al.}(2019)El-Badry, Rix, Tian, Duch\^ene, \&
  Moe}]{ElBadry2019Twins}
El-Badry, K., Rix, H.-W., Tian, H., Duch\^ene, G., \& Moe, M. 2019,
  \bibinfo{title}{Discovery of an equal-mass ‘twin’ binary population
  reaching 1000 + au separations,} \mnras, 489, 5822–5857,
  \dodoi{10.1093/mnras/stz2480}

% type= article
\bibitem[{D.~K. {Feuillet} {et~al.}(2019){Feuillet}, {Frankel}, {Lind},
  {Frinchaboy}, {García-Hernández}, {Lane}, {Nitschelm}, \&
  {Roman-Lopes}}]{Feuillet2019}
{Feuillet}, D.~K., {Frankel}, N., {Lind}, K., {et~al.} 2019,
  \bibinfo{title}{{Spatial variations in the Milky Way disc metallicity–age
  relation},} \mnras, 489, 1742–1752, \dodoi{10.1093/mnras/stz2221}

% type= article
\bibitem[{D.~K. {Feuillet} {et~al.}(2018){Feuillet}, {Bovy}, {Holtzman},
  {Weinberg}, {García-Hernández}, {Hearty}, {Majewski}, {Roman-Lopes},
  {Rybizki}, \& {Zamora}}]{Feuillet2018}
{Feuillet}, D.~K., {Bovy}, J., {Holtzman}, J., {et~al.} 2018,
  \bibinfo{title}{{Age-resolved chemistry of red giants in the solar
  neighbourhood},} \mnras, 477, 2326–2348, \dodoi{10.1093/mnras/sty779}

% type= article
\bibitem[{D. Grether \& C.~H. Lineweaver(2007)Grether \&
  Lineweaver}]{Grether2007}
Grether, D., \& Lineweaver, C.~H. 2007, \bibinfo{title}{The Metallicity of
  Stars with Close Companions,} \apj, 669, 1220–1234, \dodoi{10.1086/521714}

% type= article
\bibitem[{C. Hamilton(2022)Hamilton}]{Hamilton2022}
Hamilton, C. 2022, \bibinfo{title}{On the Phase-mixed Eccentricity and
  Inclination Distributions of Wide Binaries in the Galaxy,} \apjl, 929, L29,
  \dodoi{10.3847/2041-8213/ac6600}

% type= article
\bibitem[{C. Hamilton \& S. Modak(2024)Hamilton \& Modak}]{Hamilton2024}
Hamilton, C., \& Modak, S. 2024, \bibinfo{title}{Eccentricity dynamics of wide
  binaries – II. The effect of stellar encounters and constraints on
  formation channels,} \mnras, 532, 2425–2440, \dodoi{10.1093/mnras/stae1654}

% type= article
\bibitem[{K. Hawkins {et~al.}(2020)Hawkins, Lucey, Ting, Ji, Katzberg,
  Thompson, El-Badry, Teske, Nelson, \& Carrillo}]{Hawkins2020}
Hawkins, K., Lucey, M., Ting, Y.-S., {et~al.} 2020, \bibinfo{title}{Identical
  or fraternal twins? The chemical homogeneity of wide binaries from
  <i>Gaia</i> DR2,} \mnras, 492, 1164–1179, \dodoi{10.1093/mnras/stz3132}

% type= article
\bibitem[{M. Haywood {et~al.}(2013)Haywood, Di~Matteo, Lehnert, Katz, \&
  Gómez}]{Haywood2013}
Haywood, M., Di~Matteo, P., Lehnert, M.~D., Katz, D., \& Gómez, A. 2013,
  \bibinfo{title}{The age structure of stellar populations in the solar
  vicinity: Clues of a two-phase formation history of the Milky Way disk,}
  \aap, 560, A109, \dodoi{10.1051/0004-6361/201321397}

% type= article
\bibitem[{D.~C. Heggie(1975)Heggie}]{Heggie1975}
Heggie, D.~C. 1975, \bibinfo{title}{Binary Evolution in Stellar Dynamics,}
  \mnras, 173, 729–787, \dodoi{10.1093/mnras/173.3.729}

% type= article
\bibitem[{D. Huber {et~al.}(2016)Huber, Bryson, Haas, Barclay, Barentsen,
  Howell, Sharma, Stello, \& Thompson}]{Huber2016}
Huber, D., Bryson, S.~T., Haas, M.~R., {et~al.} 2016, \bibinfo{title}{THE K2
  ECLIPTIC PLANE INPUT CATALOG (EPIC) AND STELLAR CLASSIFICATIONS OF 138, 600
  TARGETS IN CAMPAIGNS 1–8,} \apjs, 224, 2, \dodoi{10.3847/0067-0049/224/1/2}

% type= article
\bibitem[{H.-C. Hwang {et~al.}(2022{\natexlab{a}})Hwang, El-Badry, Rix,
  Hamilton, Ting, \& Zakamska}]{Hwang2022Twins}
Hwang, H.-C., El-Badry, K., Rix, H.-W., {et~al.} 2022{\natexlab{a}},
  \bibinfo{title}{Wide {Twin} {Binaries} are {Extremely} {Eccentric}:
  {Evidence} of {Twin} {Binary} {Formation} in {Circumbinary} {Disks},} \apjl,
  933, L32, \dodoi{10.3847/2041-8213/ac7c70}

% type= article
\bibitem[{H.-C. {Hwang} {et~al.}(2021){Hwang}, {Ting}, {Schlaufman},
  {Zakamska}, \& {Wyse}}]{Hwang2021}
{Hwang}, H.-C., {Ting}, Y.-S., {Schlaufman}, K.~C., {Zakamska}, N.~L., \&
  {Wyse}, R. F.~G. 2021, \bibinfo{title}{{The non-monotonic, strong metallicity
  dependence of the wide-binary fraction},} \mnras, 501, 4329,
  \dodoi{10.1093/mnras/staa3854}

% type= article
\bibitem[{H.-C. Hwang {et~al.}(2022{\natexlab{b}})Hwang, Ting, Conroy,
  Zakamska, El-Badry, Cargile, Zaritsky, Chandra, Han, Speagle, \&
  Bonaca}]{Hwang2022}
Hwang, H.-C., Ting, Y.-S., Conroy, C., {et~al.} 2022{\natexlab{b}},
  \bibinfo{title}{Wide binaries from the H3 survey: the thick disc and halo
  have similar wide binary fractions,} \mnras, 513, 754–767,
  \dodoi{10.1093/mnras/stac650}

% type= article
\bibitem[{Y.-F. Jiang \& S. Tremaine(2010)Jiang \& Tremaine}]{Jiang2010}
Jiang, Y.-F., \& Tremaine, S. 2010, \bibinfo{title}{The evolution of wide
  binary stars,} \mnras, 401, 977–994,
  \dodoi{10.1111/j.1365-2966.2009.15744.x}

% type= article
\bibitem[{N.~A. Kaib \& S.~N. Raymond(2014)Kaib \& Raymond}]{Kaib2014}
Kaib, N.~A., \& Raymond, S.~N. 2014, \bibinfo{title}{VERY WIDE BINARY STARS AS
  THE PRIMARY SOURCE OF STELLAR COLLISIONS IN THE GALAXY,} \apj, 782, 60,
  \dodoi{10.1088/0004-637x/782/2/60}

% type= article
\bibitem[{N.~A. Kaib {et~al.}(2013)Kaib, Raymond, \& Duncan}]{Kaib2013}
Kaib, N.~A., Raymond, S.~N., \& Duncan, M. 2013, \bibinfo{title}{Planetary
  system disruption by Galactic perturbations to wide binary stars,} Nature,
  493, 381–384, \dodoi{10.1038/nature11780}

% type= article
\bibitem[{M.~B.~N. Kouwenhoven {et~al.}(2010)Kouwenhoven, Goodwin, Parker,
  Davies, Malmberg, \& Kroupa}]{Kouwenhoven2010}
Kouwenhoven, M. B.~N., Goodwin, S.~P., Parker, R.~J., {et~al.} 2010,
  \bibinfo{title}{The formation of very wide binaries during the star cluster
  dissolution phase,} \mnras, 404, 1835,
  \dodoi{10.1111/j.1365-2966.2010.16399.x}

% type= article
\bibitem[{A.~L. Kraus \& L.~A. Hillenbrand(2009)Kraus \&
  Hillenbrand}]{Kraus2009}
Kraus, A.~L., \& Hillenbrand, L.~A. 2009, \bibinfo{title}{THE COEVALITY OF
  YOUNG BINARY SYSTEMS,} \apj, 704, 531–547,
  \dodoi{10.1088/0004-637x/704/1/531}

% type= article
\bibitem[{N. Lodieu {et~al.}(2025)Lodieu, Pérez~Garrido, Zhang, Martín,
  Rebolo~López, Pérez-Toledo, Clavero, \& Nespral}]{Lodieu2025}
Lodieu, N., Pérez~Garrido, A., Zhang, J.-Y., {et~al.} 2025,
  \bibinfo{title}{The wide binary frequency of metal-poor stars,} \aap, 694,
  A129, \dodoi{10.1051/0004-6361/202450311}

% type= article
\bibitem[{Y.~L. {Lu} {et~al.}(2022){Lu}, {Ness}, {Buck}, {Zinn}, \&
  {Johnston}}]{Lu2022}
{Lu}, Y.~L., {Ness}, M.~K., {Buck}, T., {Zinn}, J.~C., \& {Johnston}, K.~V.
  2022, \bibinfo{title}{{Similarities behind the high- and low-$\alpha$ disc:
  small intrinsic abundance scatter and migrating stars},} \mnras, 512, 2890,
  \dodoi{10.1093/mnras/stac610}

% type= article
\bibitem[{A.-L. Luo {et~al.}(2015)Luo, Zhao, Zhao, Deng, Liu, Jing, Wang,
  Zhang, Shi, Cui, Chu, Li, Bai, Wu, Cai, Cao, Cao, Carlin, Chen, Chen, Chen,
  Chen, Chen, Chen, Chen, Christlieb, Chu, Cui, Dong, Du, Fan, Feng, Fu, Gao,
  Gong, Gu, Guo, Han, He, Hou, Hou, Hou, Hu, Hu, Hu, Huo, Jia, Jiang, Jiang,
  Jiang, Jin, Kong, Kong, Lei, Li, Li, Li, Li, Li, Li, Li, Li, Li, Li, Li, Li,
  Liang, Lin, Liu, Liu, Liu, Liu, Lu, Luo, Mao, Newberg, Ni, Qi, Qi, Shen, Shi,
  Song, Song, Su, Su, Tang, Tao, Tian, Wang, Wang, Wang, Wang, Wang, Wang,
  Wang, Wang, Wang, Wang, Wang, Wang, Wang, Wang, Wang, Wang, Wang, Wang, Wang,
  Wang, Wei, Wei, Wu, Wu, Wu, Wu, Xing, Xu, Xu, Xu, Yan, Yang, Yang, Yang,
  Yang, Yao, Yu, Yuan, Yuan, Yuan, Yuan, Zhai, Zhang, Zhang, Zhang, Zhang,
  Zhang, Zhang, Zhang, Zhang, Zhao, Zhou, Zhou, Zhu, Zhu, Zou, \&
  Zuo}]{Luo2015}
Luo, A.-L., Zhao, Y.-H., Zhao, G., {et~al.} 2015, \bibinfo{title}{The first
  data release (DR1) of the LAMOST regular survey,} RAA, 15, 1095–1124,
  \dodoi{10.1088/1674-4527/15/8/002}

% type= article
\bibitem[{S. Lépine \& B. Bongiorno(2007)Lépine \& Bongiorno}]{Lpine2007}
Lépine, S., \& Bongiorno, B. 2007, \bibinfo{title}{New Distant Companions to
  Known Nearby Stars. II. Faint Companions of<i>Hipparcos</i>Stars and the
  Frequency of Wide Binary Systems,} \aj, 133, 889–905,
  \dodoi{10.1086/510333}

% type= article
\bibitem[{M.~N. Machida {et~al.}(2009)Machida, Omukai, Matsumoto, \&
  Inutsuka}]{Machida2009}
Machida, M.~N., Omukai, K., Matsumoto, T., \& Inutsuka, S.-i. 2009,
  \bibinfo{title}{Binary formation with different metallicities: dependence on
  initial conditions,} \mnras, 399, 1255–1263,
  \dodoi{10.1111/j.1365-2966.2009.15394.x}

% type= article
\bibitem[{V.~V. Makarov {et~al.}(2008)Makarov, Zacharias, \&
  Hennessy}]{Makarov2008}
Makarov, V.~V., Zacharias, N., \& Hennessy, G.~S. 2008, \bibinfo{title}{Common
  Proper Motion Companions to Nearby Stars: Ages and Evolution,} \apj, 687,
  566–578, \dodoi{10.1086/591638}

% type= article
\bibitem[{S. Modak \& C. Hamilton(2023)Modak \& Hamilton}]{Modak2023}
Modak, S., \& Hamilton, C. 2023, \bibinfo{title}{Eccentricity dynamics of wide
  binaries – I. The effect of Galactic tides,} \mnras, 524, 3102–3115,
  \dodoi{10.1093/mnras/stad2073}

% type= article
\bibitem[{M. Moe \& R. Di~Stefano(2017)Moe \& Di~Stefano}]{Moe2017}
Moe, M., \& Di~Stefano, R. 2017, \bibinfo{title}{Mind Your Ps and Qs: The
  Interrelation between Period (P) and Mass-ratio (Q) Distributions of Binary
  Stars,} \apjs, 230, 15, \dodoi{10.3847/1538-4365/aa6fb6}

% type= article
\bibitem[{M. Moe {et~al.}(2019)Moe, Kratter, \& Badenes}]{Moe2019}
Moe, M., Kratter, K.~M., \& Badenes, C. 2019, \bibinfo{title}{The Close Binary
  Fraction of Solar-type Stars Is Strongly Anticorrelated with Metallicity,}
  \apj, 875, 61, \dodoi{10.3847/1538-4357/ab0d88}

% type= article
\bibitem[{Z. {Niu} {et~al.}(2022){Niu}, {Yuan}, {Wang}, \& {Liu}}]{Niu2022}
{Niu}, Z., {Yuan}, H., {Wang}, Y., \& {Liu}, J. 2022, \bibinfo{title}{{Not That
  Simple: The Metallicity Dependence of the Wide Binary Fraction Changes with
  Separation and Stellar Mass},} \apj, 931, 124,
  \dodoi{10.3847/1538-4357/ac6c84}

% type= article
\bibitem[{T. Prusti {et~al.}(2016)Prusti, de~Bruijne, Brown, Vallenari,
  Babusiaux, Bailer-Jones, Bastian, Biermann, Evans, Eyer, Jansen, Jordi,
  Klioner, Lammers, Lindegren, Luri, Mignard, Milligan, Panem, Poinsignon,
  Pourbaix, Randich, Sarri, Sartoretti, Siddiqui, Soubiran, Valette, van
  Leeuwen, Walton, Aerts, Arenou, Cropper, Drimmel, Høg, Katz, Lattanzi,
  O’Mullane, Grebel, Holland, Huc, Passot, Bramante, Cacciari, Castañeda,
  Chaoul, Cheek, De~Angeli, Fabricius, Guerra, Hernández,
  Jean-Antoine-Piccolo, Masana, Messineo, Mowlavi, Nienartowicz,
  Ordóñez-Blanco, Panuzzo, Portell, Richards, Riello, Seabroke, Tanga,
  Thévenin, Torra, Els, Gracia-Abril, Comoretto, Garcia-Reinaldos, Lock,
  Mercier, Altmann, Andrae, Astraatmadja, Bellas-Velidis, Benson, Berthier,
  Blomme, Busso, Carry, Cellino, Clementini, Cowell, Creevey, Cuypers,
  Davidson, De~Ridder, de~Torres, Delchambre, Dell’Oro, Ducourant, Frémat,
  García-Torres, Gosset, Halbwachs, Hambly, Harrison, Hauser, Hestroffer,
  Hodgkin, Huckle, Hutton, Jasniewicz, Jordan, Kontizas, Korn, Lanzafame,
  Manteiga, Moitinho, Muinonen, Osinde, Pancino, Pauwels, Petit, Recio-Blanco,
  Robin, Sarro, Siopis, Smith, Smith, Sozzetti, Thuillot, van Reeven, Viala,
  Abbas, Abreu~Aramburu, Accart, Aguado, Allan, Allasia, Altavilla, Álvarez,
  Alves, Anderson, Andrei, Anglada~Varela, Antiche, Antoja, Antón, Arcay,
  Atzei, Ayache, Bach, Baker, Balaguer-Núñez, Barache, Barata, Barbier,
  Barblan, Baroni, Barrado~y Navascués, Barros, Barstow, Becciani, Bellazzini,
  Bellei, Bello~García, Belokurov, Bendjoya, Berihuete, Bianchi, Bienaymé,
  Billebaud, Blagorodnova, Blanco-Cuaresma, Boch, Bombrun, Borrachero,
  Bouquillon, Bourda, Bouy, Bragaglia, Breddels, Brouillet, Br\"{u}semeister,
  Bucciarelli, Budnik, Burgess, Burgon, Burlacu, Busonero, Buzzi, Caffau,
  Cambras, Campbell, Cancelliere, Cantat-Gaudin, Carlucci, Carrasco,
  Castellani, Charlot, Charnas, Charvet, Chassat, Chiavassa, Clotet, Cocozza,
  Collins, Collins, Costigan, Crifo, Cross, Crosta, Crowley, Dafonte, Damerdji,
  Dapergolas, David, David, De~Cat, de~Felice, de~Laverny, De~Luise, De~March,
  de~Martino, de~Souza, Debosscher, del Pozo, Delbo, Delgado, Delgado,
  di~Marco, Di~Matteo, Diakite, Distefano, Dolding, Dos~Anjos, Drazinos,
  Durán, Dzigan, Ecale, Edvardsson, Enke, Erdmann, Escolar, Espina, Evans,
  Eynard~Bontemps, Fabre, Fabrizio, Faigler, Falcão, Farràs~Casas, Faye,
  Federici, Fedorets, Fernández-Hernández, Fernique, Fienga, Figueras,
  Filippi, Findeisen, Fonti, Fouesneau, Fraile, Fraser, Fuchs, Furnell, Gai,
  Galleti, Galluccio, Garabato, García-Sedano, Garé, Garofalo, Garralda,
  Gavras, Gerssen, Geyer, Gilmore, Girona, Giuffrida, Gomes, González-Marcos,
  González-Núñez, González-Vidal, Granvik, Guerrier, Guillout, Guiraud,
  Gúrpide, Gutiérrez-Sánchez, Guy, Haigron, Hatzidimitriou, Haywood, Heiter,
  Helmi, Hobbs, Hofmann, Holl, Holland, Hunt, Hypki, Icardi, Irwin, Jevardat~de
  Fombelle, Jofré, Jonker, Jorissen, Julbe, Karampelas, Kochoska, Kohley,
  Kolenberg, Kontizas, Koposov, Kordopatis, Koubsky, Kowalczyk, Krone-Martins,
  Kudryashova, Kull, Bachchan, Lacoste-Seris, Lanza, Lavigne,
  Le~Poncin-Lafitte, Lebreton, Lebzelter, Leccia, Leclerc, Lecoeur-Taibi,
  Lemaitre, Lenhardt, Leroux, Liao, Licata, Lindstrøm, Lister, Livanou, Lobel,
  L\"{o}ffler, López, Lopez-Lozano, Lorenz, Loureiro, MacDonald,
  Magalhães~Fernandes, Managau, Mann, Mantelet, Marchal, Marchant, Marconi,
  Marie, Marinoni, Marrese, Marschalkó, Marshall, Martín-Fleitas, Martino,
  Mary, Matijevič, Mazeh, McMillan, Messina, Mestre, Michalik, Millar,
  Miranda, Molina, Molinaro, Molinaro, Molnár, Moniez, Montegriffo, Monteiro,
  Mor, Mora, Morbidelli, Morel, Morgenthaler, Morley, Morris, Mulone, Muraveva,
  Musella, Narbonne, Nelemans, Nicastro, Noval, Ordénovic, Ordieres-Meré,
  Osborne, Pagani, Pagano, Pailler, Palacin, Palaversa, Parsons, Paulsen,
  Pecoraro, Pedrosa, Pentik\"{a}inen, Pereira, Pichon, Piersimoni, Pineau,
  Plachy, Plum, Poujoulet, Prša, Pulone, Ragaini, Rago, Rambaux, Ramos-Lerate,
  Ranalli, Rauw, Read, Regibo, Renk, Reylé, Ribeiro, Rimoldini, Ripepi, Riva,
  Rixon, Roelens, Romero-Gómez, Rowell, Royer, Rudolph, Ruiz-Dern, Sadowski,
  Sagristà~Sellés, Sahlmann, Salgado, Salguero, Sarasso, Savietto, Schnorhk,
  Schultheis, Sciacca, Segol, Segovia, Segransan, Serpell, Shih, Smareglia,
  Smart, Smith, Solano, Solitro, Sordo, Soria~Nieto, Souchay, Spagna, Spoto,
  Stampa, Steele, Steidelm\"{u}ller, Stephenson, Stoev, Suess, S\"{u}veges,
  Surdej, Szabados, Szegedi-Elek, Tapiador, Taris, Tauran, Taylor, Teixeira,
  Terrett, Tingley, Trager, Turon, Ulla, Utrilla, Valentini, van Elteren,
  Van~Hemelryck, van Leeuwen, Varadi, Vecchiato, Veljanoski, Via, Vicente,
  Vogt, Voss, Votruba, Voutsinas, Walmsley, Weiler, Weingrill, Werner, Wevers,
  Whitehead, Wyrzykowski, Yoldas, Žerjal, Zucker, Zurbach, Zwitter, Alecu,
  Allen, Allende~Prieto, Amorim, Anglada-Escudé, Arsenijevic, Azaz, Balm,
  Beck, Bernstein, Bigot, Bijaoui, Blasco, Bonfigli, Bono, Boudreault, Bressan,
  Brown, Brunet, Bunclark, Buonanno, Butkevich, Carret, Carrion, Chemin,
  Chéreau, Corcione, Darmigny, de~Boer, de~Teodoro, de~Zeeuw, Delle~Luche,
  Domingues, Dubath, Fodor, Frézouls, Fries, Fustes, Fyfe, Gallardo, Gallegos,
  Gardiol, Gebran, Gomboc, Gómez, Grux, Gueguen, Heyrovsky, Hoar, Iannicola,
  Isasi~Parache, Janotto, Joliet, Jonckheere, Keil, Kim, Klagyivik, Klar,
  Knude, Kochukhov, Kolka, Kos, Kutka, Lainey, LeBouquin, Liu, Loreggia,
  Makarov, Marseille, Martayan, Martinez-Rubi, Massart, Meynadier, Mignot,
  Munari, Nguyen, Nordlander, Ocvirk, O’Flaherty, Olias~Sanz, Ortiz, Osorio,
  Oszkiewicz, Ouzounis, Palmer, Park, Pasquato, Peltzer, Peralta, Péturaud,
  Pieniluoma, Pigozzi, Poels, Prat, Prod’homme, Raison, Rebordao, Risquez,
  Rocca-Volmerange, Rosen, Ruiz-Fuertes, Russo, Sembay, Serraller~Vizcaino,
  Short, Siebert, Silva, Sinachopoulos, Slezak, Soffel, Sosnowska, Straižys,
  ter Linden, Terrell, Theil, Tiede, Troisi, Tsalmantza, Tur, Vaccari, Vachier,
  Valles, Van~Hamme, Veltz, Virtanen, Wallut, Wichmann, Wilkinson, Ziaeepour,
  \& Zschocke}]{Gaia2016}
Prusti, T., de~Bruijne, J. H.~J., Brown, A. G.~A., {et~al.} 2016,
  \bibinfo{title}{The Gaia mission,} \aap, 595, A1,
  \dodoi{10.1051/0004-6361/201629272}

% type= misc
\bibitem[{T. Qiu {et~al.}(2026)Qiu, Wang, Koposov, Li, Sandford, Li, Najita,
  Aguilar, Ahlen, Bianchi, Brooks, Rosell, Claybaugh, Cuceu, de~la Macorra,
  Dey, Doel, Forero-Romero, Gaztañaga, Gontcho, Gutierrez, Juneau, Kehoe,
  Landriau, Guillou, Meisner, Miquel, Paillas, Percival, Prada, Pérez-Ràfols,
  Ravoux, Rossi, Sanchez, Saulder, Schlafly, Schlegel, Schubnell, Silber,
  Siudek, Tarlé, \& Weaver}]{Qiu2026}
Qiu, T., Wang, W., Koposov, S.~E., {et~al.} 2026, The Binary Fraction of Milky
  Way Field Stars in DESI: Dependence on Chemical Abundance, arXiv,
  \dodoi{10.48550/ARXIV.2610.11102}

% type= article
\bibitem[{D. Raghavan {et~al.}(2010)Raghavan, McAlister, Henry, Latham, Marcy,
  Mason, Gies, White, \& ten Brummelaar}]{Raghavan2010}
Raghavan, D., McAlister, H.~A., Henry, T.~J., {et~al.} 2010, \bibinfo{title}{A
  SURVEY OF STELLAR FAMILIES: MULTIPLICITY OF SOLAR-TYPE STARS,} \apjs, 190,
  1–42, \dodoi{10.1088/0067-0049/190/1/1}

% type= article
\bibitem[{E.~D. Ramirez \& M.~R. Buckley(2023)Ramirez \& Buckley}]{Ramirez2023}
Ramirez, E.~D., \& Buckley, M.~R. 2023, \bibinfo{title}{Constraining dark
  matter substructure with Gaia wide binaries,} \mnras, 525, 5813–5830,
  \dodoi{10.1093/mnras/stad2583}

% type= article
\bibitem[{R. {Schönrich} \& M. {Aumer}(2017){Schönrich} \&
  {Aumer}}]{Schonrich2017}
{Schönrich}, R., \& {Aumer}, M. 2017, \bibinfo{title}{{Assessing distances and
  consistency of kinematics in Gaia/TGAS},} \mnras, 472, 3979–3998,
  \dodoi{10.1093/mnras/stx2189}

% type= article
\bibitem[{C. Shariat {et~al.}(2026)Shariat, El-Badry, \&
  Bhattacharjee}]{Shariat2026}
Shariat, C., El-Badry, K., \& Bhattacharjee, S. 2026, \bibinfo{title}{How
  precisely can we measure the ages of subgiant and giant stars?} The Open
  Journal of Astrophysics, 9, \dodoi{10.33232/001c.162534}

% type= article
\bibitem[{C. Shariat {et~al.}(2025{\natexlab{a}})Shariat, El-Badry, \&
  Naoz}]{Shariat2025Triples}
Shariat, C., El-Badry, K., \& Naoz, S. 2025{\natexlab{a}},
  \bibinfo{title}{10,000 Resolved Triples from Gaia: Empirical Constraints on
  Triple Star Populations,} \pasp, 137, 094201,
  \dodoi{10.1088/1538-3873/adfb30}

% type= article
\bibitem[{C. Shariat {et~al.}(2025{\natexlab{b}})Shariat, El-Badry, Gennaro,
  Ding~丁, Simon, Avila, Calamida, Cassisi, Correnti, Weisz, Geha, Kirby,
  Brown, Ricotti, McQuinn, Kallivayalil, Gilbert, Pacifici, Guhathakurta,
  Crnojević, Boyer, Beaton, Chandra, Cohen, Renzini, Savino, \&
  Tollerud}]{Shariat2025}
Shariat, C., El-Badry, K., Gennaro, M., {et~al.} 2025{\natexlab{b}},
  \bibinfo{title}{Wide Binaries in an Ultra-faint Dwarf Galaxy: Discovery,
  Population Modeling, and A Nail in the Coffin of Primordial black hole Dark
  Matter,} \pasp, 137, 104103, \dodoi{10.1088/1538-3873/ae11d2}

% type= article
\bibitem[{T. {Sun} {et~al.}(2023){Sun}, {Chen}, {Bi}, {Ge}, {Xiang}, \&
  {Wu}}]{Sun2023}
{Sun}, T., {Chen}, X., {Bi}, S., {et~al.} 2023, \bibinfo{title}{{Characterizing
  abundance–age relations of GALAH stars using oxygen-enhanced stellar
  models},} \mnras, 523, 1199–1208, \dodoi{10.1093/mnras/stad1499}

% type= article
\bibitem[{H. {Tian} {et~al.}(2019){Tian}, {Liu}, {Xu}, \& {Xue}}]{Tian2019}
{Tian}, H., {Liu}, C., {Xu}, Y., \& {Xue}, X. 2019, \bibinfo{title}{{Rotating
  Halo Traced by the K-giant Stars from LAMOST and Gaia},} \apj, 871, 184,
  \dodoi{10.3847/1538-4357/aaf6e8}

% type= article
\bibitem[{H.-J. {Tian} {et~al.}(2020){Tian}, {El-Badry}, {Rix}, \&
  {Gould}}]{Tian2019a}
{Tian}, H.-J., {El-Badry}, K., {Rix}, H.-W., \& {Gould}, A. 2020,
  \bibinfo{title}{{The Separation Distribution of Ultrawide Binaries across
  Galactic Populations},} \apjs, 246, 4, \dodoi{10.3847/1538-4365/ab54c4}

% type= article
\bibitem[{H.-J. {Tian} {et~al.}(2015){Tian}, {Liu}, {Carlin}, {Zhao}, {Chen},
  {Wu}, {Li}, {Hou}, \& {Zhang}}]{Tian2015}
{Tian}, H.-J., {Liu}, C., {Carlin}, J.~L., {et~al.} 2015, \bibinfo{title}{{THE
  STELLAR KINEMATICS IN THE SOLAR NEIGHBORHOOD FROM LAMOST DATA},} \apj, 809,
  145, \dodoi{10.1088/0004-637X/809/2/145}

% type= article
\bibitem[{Y.-S. Ting {et~al.}(2019)Ting, Conroy, Rix, \& Cargile}]{Ting2019}
Ting, Y.-S., Conroy, C., Rix, H.-W., \& Cargile, P. 2019, \bibinfo{title}{The
  Payne: Self-consistent ab initio Fitting of Stellar Spectra,} \apj, 879, 69,
  \dodoi{10.3847/1538-4357/ab2331}

% type= article
\bibitem[{A. Tokovinin(2014)Tokovinin}]{Tokovinin2014}
Tokovinin, A. 2014, \bibinfo{title}{FROM BINARIES TO MULTIPLES. II.
  HIERARCHICAL MULTIPLICITY OF F AND G DWARFS,} \aj, 147, 87,
  \dodoi{10.1088/0004-6256/147/4/87}

% type= article
\bibitem[{J.-H. Wang {et~al.}(2025)Wang, Xiang, Zhang, Xie, Ge, Zhang, Mou, \&
  Liu}]{Wang2025}
Wang, J.-H., Xiang, M., Zhang, M., {et~al.} 2025, \bibinfo{title}{Spectroscopic
  Ages for 4 Million Main-sequence Dwarf Stars from LAMOST DR10 Estimated with
  a Data-driven Approach,} \apjs, 280, 13, \dodoi{10.3847/1538-4365/aded16}

% type= article
\bibitem[{M.~D. Weinberg {et~al.}(1987)Weinberg, Shapiro, \&
  Wasserman}]{Weinberg1987}
Weinberg, M.~D., Shapiro, S.~L., \& Wasserman, I. 1987, \bibinfo{title}{The
  dynamical fate of wide binaries in the solar neighborhood,} \apj, 312, 367,
  \dodoi{10.1086/164883}

% type= article
\bibitem[{M. {Xiang} \& H.-W. {Rix}(2022){Xiang} \& {Rix}}]{Xiang2022}
{Xiang}, M., \& {Rix}, H.-W. 2022, \bibinfo{title}{{A time-resolved picture of
  our Milky Way’s early formation history},} Nature, 603, 599,
  \dodoi{10.1038/s41586-022-04496-5}

% type= article
\bibitem[{M. Xiang {et~al.}(2024)Xiang, Rix, Yang, Liu, Huang, \&
  Frankel}]{Xiang2024}
Xiang, M., Rix, H.-W., Yang, H., {et~al.} 2024, \bibinfo{title}{The formation
  and survival of the Milky Way’s oldest stellar disk,} Nature Astronomy, 9,
  101–110, \dodoi{10.1038/s41550-024-02382-w}

% type= article
\bibitem[{M. {Xiang} {et~al.}(2018){Xiang}, {Shi}, {Liu}, {Yuan}, {Chen},
  {Huang}, {Wang}, {Wu}, {Tian}, {Huo}, {Zhang}, \& {Zhang}}]{Xiang2018}
{Xiang}, M., {Shi}, J., {Liu}, X., {et~al.} 2018, \bibinfo{title}{{Stellar Mass
  Distribution and Star Formation History of the Galactic Disk Revealed by
  Mono-age Stellar Populations from LAMOST},} \apjs, 237, 33,
  \dodoi{10.3847/1538-4365/aad237}

% type= article
\bibitem[{M. {Xiang} {et~al.}(2019){Xiang}, {Ting}, {Rix}, {Sandford}, {Buder},
  {Lind}, {Liu}, {Shi}, \& {Zhang}}]{Xiang2019}
{Xiang}, M., {Ting}, Y.-S., {Rix}, H.-W., {et~al.} 2019,
  \bibinfo{title}{{Abundance Estimates for 16 Elements in 6 Million Stars from
  LAMOST DR5 Low-Resolution Spectra},} \apjs, 245, 34,
  \dodoi{10.3847/1538-4365/ab5364}

% type= article
\bibitem[{T.~D. Yavetz(2025)Yavetz}]{Yavetz2025}
Yavetz, T.~D. 2025, \bibinfo{title}{Disrupted Wide Binaries as Dynamical Probes
  of Galactic Structure,} \apj, 981, 124, \dodoi{10.3847/1538-4357/adaeac}

% type= article
\bibitem[{J. Yoo {et~al.}(2004)Yoo, Chanamé, \& Gould}]{Yoo2004}
Yoo, J., Chanamé, J., \& Gould, A. 2004, \bibinfo{title}{The End of the MACHO
  Era: Limits on Halo Dark Matter from Stellar Halo Wide Binaries,} \apj, 601,
  311–318, \dodoi{10.1086/380562}

% type= article
\bibitem[{M.~R. Zapatero~Osorio \& E.~L. Martín(2004)Zapatero~Osorio \&
  Martín}]{ZapateroOsorio2004}
Zapatero~Osorio, M.~R., \& Martín, E.~L. 2004, \bibinfo{title}{A CCD imaging
  search for wide metal-poor binaries,} \aap, 419, 167–180,
  \dodoi{10.1051/0004-6361:20035907}

% type= article
\bibitem[{M. {Zhang} {et~al.}(2025){Zhang}, {Xiang}, {Ting}, {Amarsi}, {Zhang},
  {Shi}, {Yuan}, {Li}, {Wang}, {Wu}, {Wu}, {Mou}, {Yan}, \& {Liu}}]{Zhang2025}
{Zhang}, M., {Xiang}, M., {Ting}, Y.-S., {et~al.} 2025,
  \bibinfo{title}{{Homogeneous Stellar Atmospheric Parameters and 22 Elemental
  Abundances for FGK Stars Derived from LAMOST Low-resolution Spectra with
  DD-PAYNE},} \apjs, 279, 5, \dodoi{10.3847/1538-4365/add016}

\end{thebibliography}

\appendix

\section{Validation of Stellar Ages and Robustness Tests}
\label{app:component_consistency}

{We perform several tests to assess the reliability of the stellar ages and the robustness of the age-based results discussed in Section~\ref{sec:age_feh}. Table~\ref{tab:age_precision} summarizes the age properties of the low-$\alpha$ sample used in our analysis. The median relative age uncertainties are 25.2\%, 15.6\%, and 16.8\% for $T_1$, $T_2$, and $T_3$, respectively. The table also gives the number of systems retained after requiring ${\rm S/N}>50$, which is used in the robustness test below.}

\begin{deluxetable}{lcccccc}[htbp]
\tablecaption{Age Properties of the Low-$\alpha$ Sample
\label{tab:age_precision}}
\tablewidth{0pt}
\tablehead{
\colhead{Sample} &
\colhead{$T_{\rm eff}$(K)} &
\colhead{$N$} &
\colhead{Median Age(Gyr)} &
\colhead{Median $\sigma_{\log {\rm Age}}$(dex)} &
\colhead{Median Relative Uncertainty (\%)} &
\colhead{$N$ (${\rm S/N}>50$)}
}
\startdata
$T_1$ & 4000--5400 & 14,194 & 4.92 & 0.109 & 25.2 & 7,241 \\
$T_2$ & 5400--6000 & 14,247 & 6.91 & 0.068 & 15.6 & 10,744 \\
$T_3$ & 6000--7500 & 13,556 & 2.48 & 0.073 & 16.8 & 11,656 \\
\enddata
\tablecomments{
{$N$ denotes the number of systems with valid ages in each $T_{\rm eff}$ subsample. The last column gives the number retained after requiring ${\rm S/N}>50$.}
}
\end{deluxetable}

\begin{figure}[!htbp]
\centering
\includegraphics[width=\textwidth]{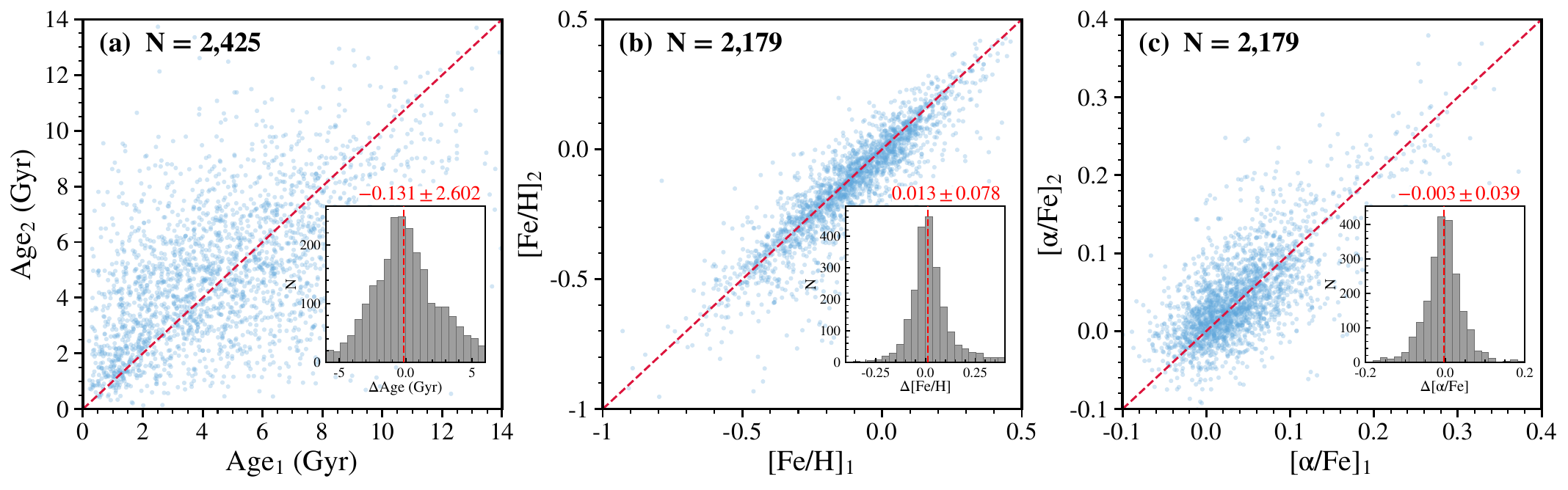}
\caption{
{Comparison of independent measurements for both components of wide binaries: (a) age for the low-$\alpha$ population, (b) [Fe/H], and (c) [$\alpha$/Fe]. Dashed lines in the main panels indicate the one-to-one relations. The inset histograms show the component differences, with red dashed lines marking the medians. Red annotations indicate the median differences and robust scatters ($\sigma_{\rm robust}=(P_{84}-P_{16})/2$).}
}
\label{fig:binary_consistency}
\end{figure}

{Wide binaries with independent spectroscopic measurements for both components provide an internal consistency check on the adopted stellar parameters. For 2,425 systems with independent age estimates, the median component difference is $-0.131$ Gyr, with a robust scatter of 2.602 Gyr. For 2,179 systems with independent abundance measurements, the median differences are 0.013 dex in [Fe/H] and $-0.003$ dex in [$\alpha$/Fe], with robust scatters of 0.078 and 0.039 dex, respectively (Figure~\ref{fig:binary_consistency}). This comparison tests the internal precision of the measurements, but does not constrain systematic uncertainties shared by both components. A similar wide-binary age-consistency test by \citet{Shariat2026} found that spectroscopic ages of subgiants can achieve fractional uncertainties of 5--10\%, whereas spectroscopic giant ages are typically less precise (25--30\%). However, these estimates concern different evolutionary stages and age estimators and are not directly comparable to the main-sequence dwarf ages adopted here.}

\begin{figure}[!htbp]
\centering
\includegraphics[width=\linewidth]{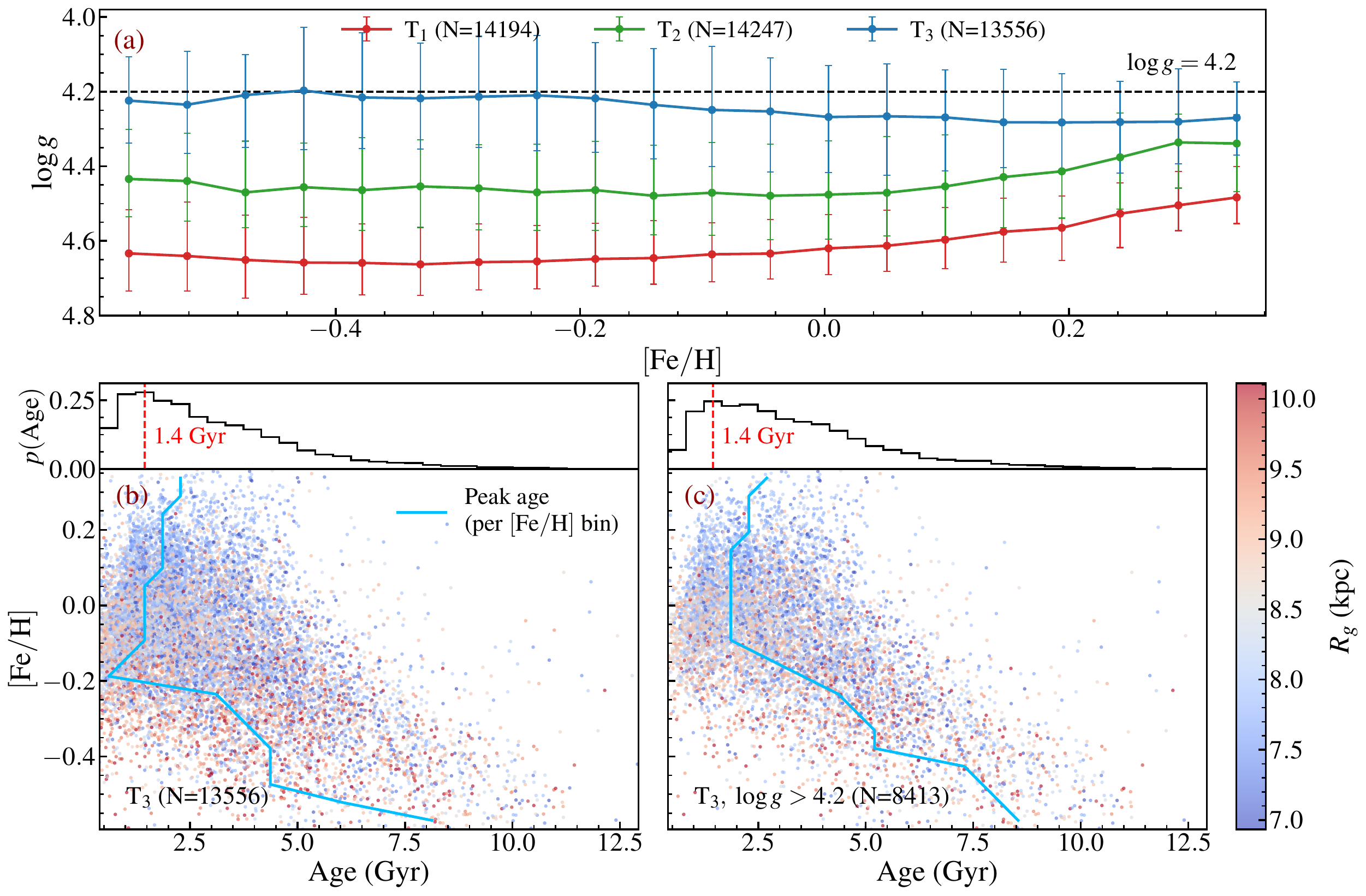}
\caption{
{Evolutionary-state test for the low-$\alpha$ sample. Panel (a) shows $\log g$ as a function of [Fe/H] for the $T_1$, $T_2$, and $T_3$ subsamples; the horizontal dashed line marks $\log g=4.2$. Panels (b) and (c) show the age--[Fe/H] distributions of the original $T_3$ sample and the restricted sample with $\log g>4.2$, respectively.}
}
\label{fig:logg_age_feh}
\end{figure}

{We next test whether the age--metallicity structure of the hot subsample is affected by stars close to the main-sequence turnoff. Adopting $\log g>4.2$ reduces the $T_3$ sample from 13,556 to 8,413 systems. The restricted sample retains the principal age--[Fe/H] structure of the original $T_3$ sample, including an age-distribution peak near 1.4 Gyr (Figure~\ref{fig:logg_age_feh}).}

\begin{figure}[!htbp]
\centering
\includegraphics[width=\linewidth]{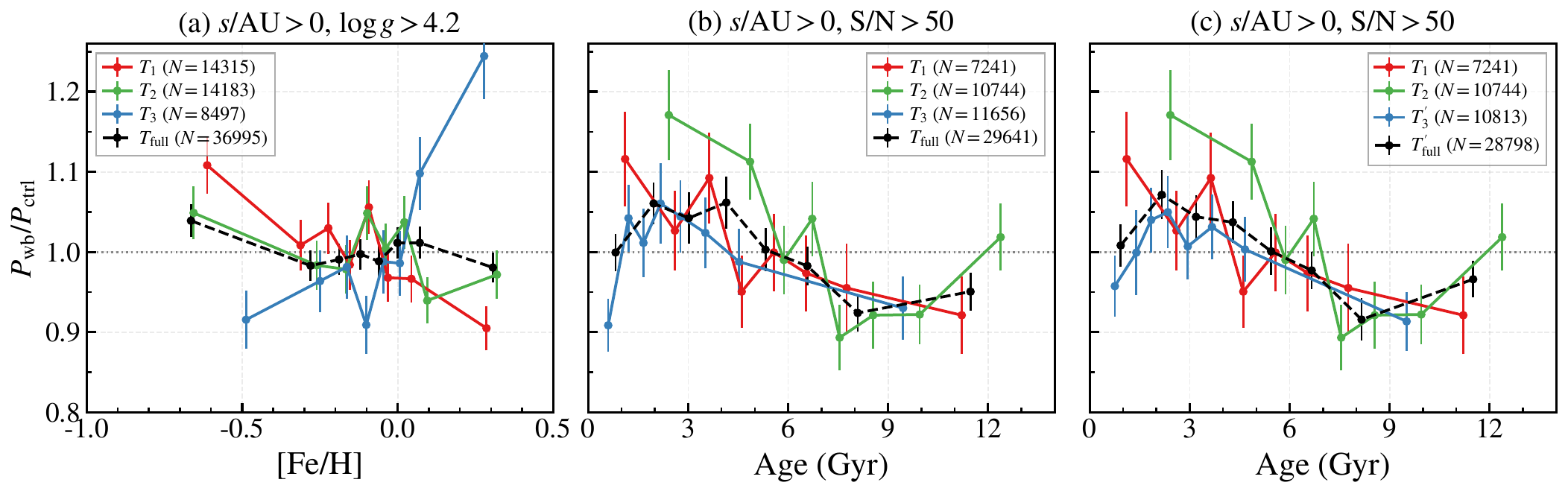}
\caption{
{Robustness tests of the relative WBF for the low-$\alpha$ population with $s>0\,{\rm AU}$. Panel (a) shows the relative WBF as a function of [Fe/H] after restricting the sample to $\log g>4.2$. Panel (b) shows the relative WBF as a function of stellar age for sample with ${\rm S/N}>50$. Panel (c) applies the same S/N cut but further restricts the temperature ranges to $6000<T_{\rm eff}<6800$ K for $T_3'$ and $4000<T_{\rm eff}<6800$ K for $T_{\rm full}'$, while keeping $T_1$ and $T_2$ unchanged.}
}
\label{fig:wbf_logg_age_robust}
\end{figure}

{Repeating the relative-WBF--[Fe/H] analysis with the same $\log g>4.2$ selection yields qualitatively similar temperature-dependent behavior for $s>0\,{\rm AU}$ (Figure~\ref{fig:wbf_logg_age_robust}(a)). Finally, we repeat the relative-WBF--age analysis using the sample with ${\rm S/N}>50$. We further test the effect of restricting the temperature ranges to $6000<T_{\rm eff}<6800$ K for $T_3'$ and $4000<T_{\rm eff}<6800$ K for $T_{\rm full}'$, while leaving $T_1$ and $T_2$ unchanged. In both cases, the resulting age dependence remains qualitatively consistent with that of the fiducial sample (Figure~\ref{fig:wbf_logg_age_robust}(b), (c)).}

\FloatBarrier
\section{Robustness to Companion-recovery Incompleteness}
\label{app:completeness}

{The twin-binary test in Section~2.4 shows that companion incompleteness increases with distance, but does not determine whether angular separation or magnitude contrast varies systematically with [Fe/H] within individual $T_{\rm eff}$ subsamples. We therefore repeat the WBF--[Fe/H] analysis for a conservative low-$\alpha$ sample restricted to $d<300$ pc, $|\Delta G|<4$, and $1000\,{\rm AU}<s<10000\,{\rm AU}$.}

\begin{figure}[!htbp]
\centering
\includegraphics[width=\linewidth]{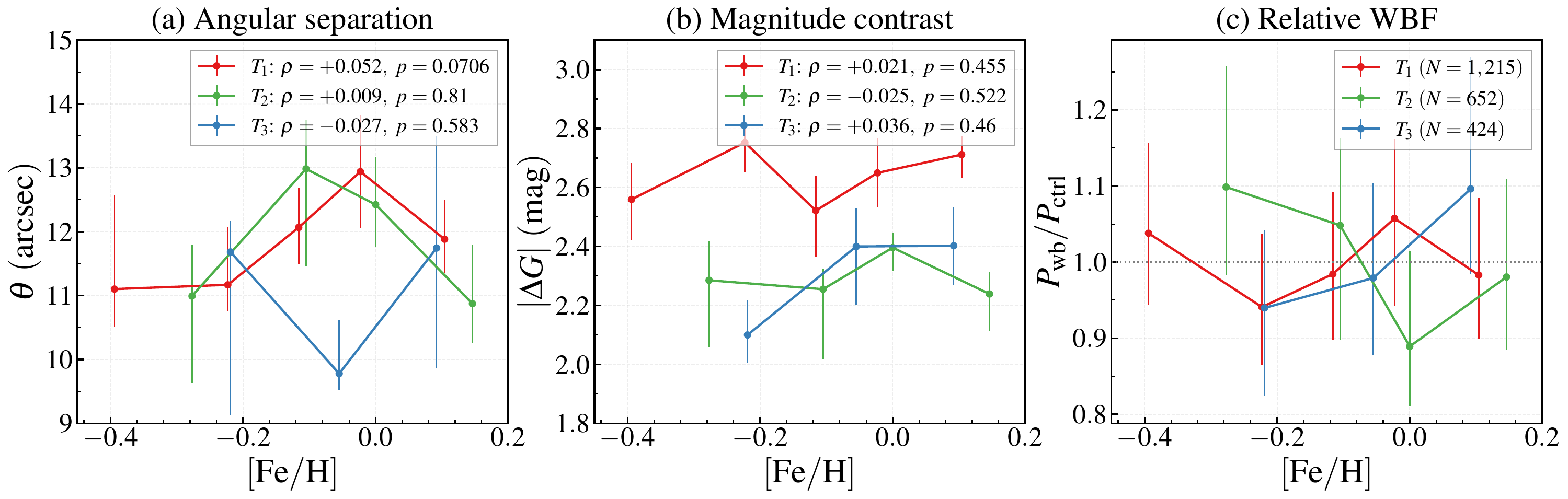}
\caption{
{Companion-recovery test for the low-$\alpha$ population with $1000\,{\rm AU}<s<10000\,{\rm AU}$, restricted to $d<300$ pc and $|\Delta G|<4$. Panels (a) and (b) show the angular separation $\theta$ and magnitude contrast $|\Delta G|$ as functions of [Fe/H], respectively; the legends give the Spearman correlation coefficients and corresponding $p$-values for each $T_{\rm eff}$ subsample. Panel (c) shows the relative WBF as a function of [Fe/H] for the same conservative sample.}
}
\label{fig:recovery_test}
\end{figure}

{Panels (a) and (b) of Figure~\ref{fig:recovery_test} show $\theta$ and $|\Delta G|$ as functions of [Fe/H] for this sample. Neither quantity exhibits a significant monotonic correlation with [Fe/H] in any $T_{\rm eff}$ subsample: $|\rho|\leq0.052$ for $\theta$ and $|\rho|\leq0.036$ for $|\Delta G|$, with all Spearman $p$-values $>0.05$. Recomputing the relative WBF for the same sample (Figure~\ref{fig:recovery_test} (c)) shows that the metal-rich increase in $T_3$ remains visible, whereas the negative $T_1$ trend is weakened. Thus, companion-recovery selection can affect the detailed amplitude of the WBF--[Fe/H] relation, particularly for $T_1$, while the temperature-dependent contrast remains present in the conservative sample.}

\FloatBarrier
\section{Comparison with Previous Work}
\label{app:hwang2021}

\begin{figure}[!htbp]
\centering
\includegraphics[width=\linewidth]{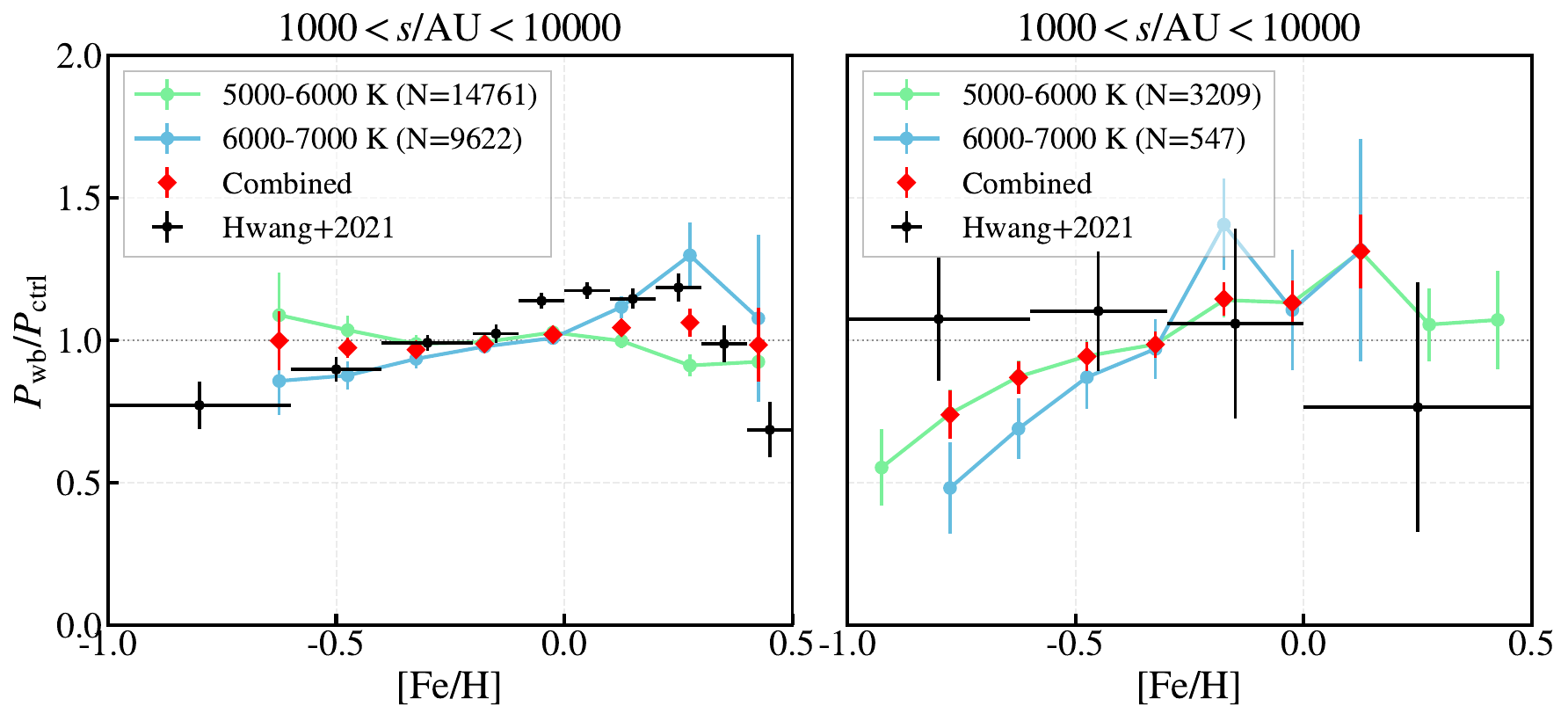}
\caption{Comparison of the WBF--[Fe/H] relations measured in this work with those of \citet{Hwang2021} for stars with $5000\,{\rm K}<T_{\rm eff}<7000\,{\rm K}$ and $1000\,{\rm AU}<s<10000\,{\rm AU}$. The left and right panels compare our low-$\alpha$ and high-$\alpha$ populations with their thin- and thick-disk samples, respectively.}
\label{fig:compare_hwang}
\end{figure}

\citet{Hwang2021} combined the Gaia DR2 wide-binary catalog with LAMOST metallicities to investigate the dependence of the WBF on [Fe/H] over the range $5000\,{\rm K}<T_{\rm eff}<7000\,{\rm K}$ and $10^3\,{\rm AU}<s<10^4\,{\rm AU}$. Using kinematic information, they divided their sample into thin- and thick-disk populations and found that both populations exhibit a non-monotonic WBF--[Fe/H] relation. To enable a direct comparison, we adopt the same $T_{\rm eff}$ and $s$ ranges. Following the $T_{\rm eff}$ distribution of their sample, we combine our $5000\,{\rm K}<T_{\rm eff}<6000\,{\rm K}$ and $6000\,{\rm K}<T_{\rm eff}<7000\,{\rm K}$ subsamples with appropriate weights. The WBF for each sample is then normalized before comparison. Figure~\ref{fig:compare_hwang} shows that the WBF--[Fe/H] relations obtained in this work are consistent with those reported by \citet{Hwang2021}.

\FloatBarrier
\section{Separation Dependence of the Relative-WBF--Age Relation}
\label{app:separation_age}

{To test whether the decline of the relative WBF with age can be attributed to cumulative dynamical disruption, we examine the age dependence in different projected-separation intervals. Because wider binaries are more weakly bound, disruption by stellar encounters, molecular clouds, and the Galactic tidal field should produce a progressively stronger age dependence toward larger separations \citep{Heggie1975,Weinberg1987,Jiang2010}.}

\begin{figure}[!htbp]
\centering
\includegraphics[width=\linewidth]{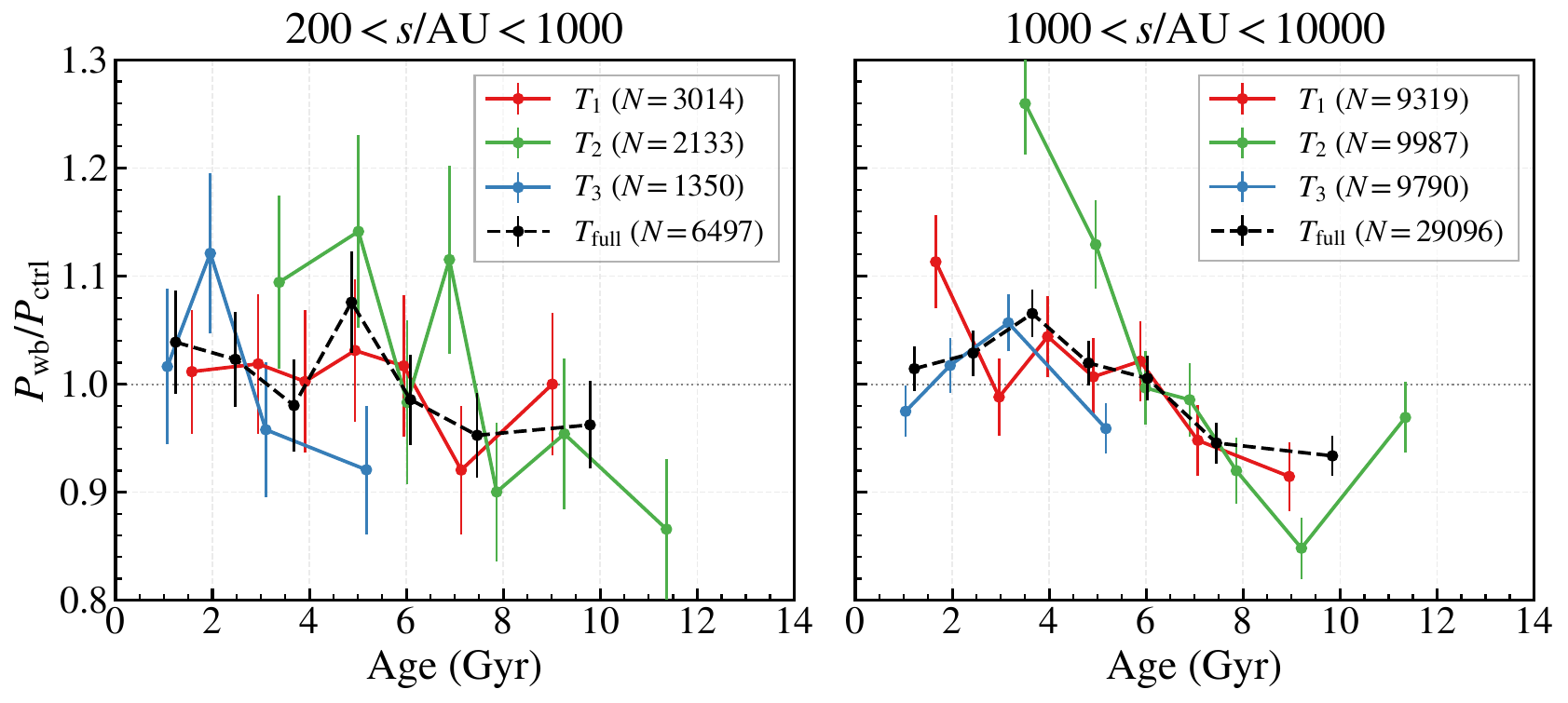}
\caption{
{Relative WBF as a function of age for the low-$\alpha$ population in the $200\,{\rm AU}<s<1000\,{\rm AU}$ and $1000\,{\rm AU}<s<10000\,{\rm AU}$. The temperature subsamples and age binning follow those adopted in the main analysis.}
}
\label{fig:wbf_age_sep}
\end{figure}

{Figure~\ref{fig:wbf_age_sep} shows the relative-WBF--age relation in the $200\,{\rm AU}<s<1000\,{\rm AU}$ and $1000\,{\rm AU}<s<10000\,{\rm AU}$ intervals. The decline does not systematically strengthen with increasing separation. This weak separation dependence is consistent with \citet{Hwang2021}, who found no significant difference in the age evolution of the WBF between binaries at $10^{2.5}$--$10^3$ AU and $10^3$--$10^4$ AU. The absence of a stronger age dependence at larger separations argues against cumulative field disruption as the dominant driver of the observed WBF evolution.}

\end{document}